\documentclass[%
aps,%
prb,%
reprint,%
amsmath,amssymb,%
floatfix,%
nofootinbib,%
superscriptaddress%
]{revtex4-2}

\usepackage[T1]{fontenc}
\usepackage[latin9]{inputenc}
\usepackage[english]{babel}
\usepackage{mathrsfs}
\usepackage{bbm}
\usepackage{graphicx}
\usepackage{xcolor}
\usepackage{physics}
\usepackage{bm}
\usepackage[export]{adjustbox}
\usepackage{tikz}

\usepackage[colorlinks,
            linkcolor=blue,    
            citecolor=blue,  
            urlcolor=blue,
 	        bookmarks=true,        
	        bookmarksopen=true,    
	        bookmarksnumbered=true,
]{hyperref}

\newcommand{\x}{{x}}

\newcommand{\imag}{{\rm Im}}

 \global\long\def\de{\delta}
 \global\long\def\Ga{\Gamma}
\global\long\def\th{\theta}
\global\long\def\th{\theta}

\global\long\def\el#1{\theta_{#1}}
\global\long\def\bell#1{\tilde\theta_{#1}}

\global\long\def\la{\lambda} 
\global\long\def\si{\sigma}
\global\long\def\vfi{\varphi}

\global\long\def\al{\alpha}
\global\long\def\be{\beta}
 \global\long\def\de{\delta}

\global\long\def\no{\nonumber}

\def\real{{\mathrm{Re}}}
\def\imag{{\mathrm{Im}}}

\global\long\def\braket#1#2{\left\langle #1|#2\right\rangle }

\def\mc{\mathcal}

\begin{document}

\title{Exact nonequilibrium steady states of boundary driven circuit with XYZ gates}

\author{Xin Zhang} 
\affiliation{Beijing National Laboratory for
  Condensed Matter Physics, Institute of Physics, Chinese Academy of
  Sciences, Beijing 100190, China}

\author{Toma\v z Prosen}
\affiliation{Faculty of Mathematics and Physics, University of Ljubljana, Jadranska 19, SI-1000 Ljubljana, Slovenia}
\affiliation{Institute of Mathematics, Physics and Mechanics, Jadranska 19, SI-1000 Ljubljana, Slovenia}

\author{Vladislav Popkov}
\affiliation{Faculty of Mathematics and Physics, University of Ljubljana, Jadranska 19, SI-1000 Ljubljana, Slovenia}
 \affiliation{Department of Physics, University of
  Wuppertal, Gaussstra\ss e 20, 42119 Wuppertal, Germany}

\begin{abstract}

We obtain the exact many-body density operator of a boundary-driven XXZ quantum circuit via a spatially inhomogeneous matrix product Ansatz.
The Ansatz has formally infinite bond-dimension and generalizes authors' previous construction \cite{2025XXZcircuit} for the XXZ interactions.
The boundary qubits are coupled to reset quantum channels that project them toward arbitrary pure target states.
We find and describe a family of relatively robust separable chiral nonequilibrium steady states (NESS), which  are elliptic analogs of spin helices for the circuit, and which
 are particularly attractive from an experimental perspective.
\end{abstract}

\maketitle

\section{Introduction} 
In today's  era of noisy  intermediate-scale quantum (NISQ) computers,  the theory of open quantum systems ~\cite{BreuerPetruccione,NielsenChuang,GardinerZoller}  is essential for building useful quantum simulations.  
In this respect exactly solvable models play a special role since they   allow to test,  control and benchmark  NISQ devices
~\cite{GateSet,GQAI_Nature22,GQAI_Science24a,GQAI_Science24b}.
 One powerful example  is a trotterized integrable XXZ quantum spin chain-realized as a brickwork quantum circuit and driven by specific Kraus dissipators at its ends,  the nonequilibrium steady state (NESS) of which can be expressed as an infinite bond-dimension matrix product ansatz (MPA) ~\cite{VanicatPRL18,2025XXZcircuit}. 
The XXZ brickwork quantum circuit  protocol 
 was implemented on an actual NISQ device ~\cite{GQAI_Nature22,GQAI_Science24b}.  
Beyond its mathematical beauty,  this setup reveals surprising and often exotic physics: for instance,  fractal-like behaviour 
of transport coefficients (Drude weights)~\cite{PRL106,LjubotinaPRL19},  insulating behavior far from equilibrium  and even 
negative differential conductance,  where increasing the driving force actually reduces the current ~\cite{Benenti,PRL107}. 

In this communication we are treating the most general,  fully anisotropic scenario when the 
principal qubit gates, constituting the XYZ-type qubit circuit,  do not satisfy the ice rule.   Namely,  we express the nonequilibrium steady state (NESS)
of the XYZ boundary driven qubit circuit in terms of MPA.  
After Ref.~\cite{2025XXZcircuit},  our  solution gives the second and more general example of the two-replica MPA that has nontrivially coupled auxiliary spaces, 
meaning that the NESS is not  factorizable in terms of a product of two  Cholesky factors,  as in all previously solved cases  ~\cite{TopicalReview}.
Analyzing the MPA,  we find a set of remarkably simple 
chiral separable states,  which,  being a circuit extensions of  spin-helix states~\cite{PopkovSHS, 2025Posske, 2022Ma,PopkovSHSkinks} 
can be named elliptic brickwork spin helices. 
The Hamiltonian (continous time) limit of elliptic spin helix eigenstates are well known~\cite{GZ1,GZ2}
even beyond one-dimensional lattices~\cite{WenWeiHo,Zheng2025}.
The XXZ counterparts of elliptic helices turned out  to be prominent in cold-atom experiments,  where the 
exceptional robustness of the helices allowed to use them for controlling and
 benchmarking purposes ~\cite{KettNature2022,2021Ketterle}.  
Likewise, we expect elliptic brickwork spin helices to be  useful benchmarking tool in potential future experiments with qubit gates not satisfying the ice rule (i.e. breaking U(1) symmetry).

\begin{figure}[tbp]
\centering
\begin{tikzpicture}[node distance=2cm, scale=0.6, transform shape]
\tikzstyle{process} = [rectangle, minimum width=1.6cm, minimum height=1cm, text centered, draw=black]
\tikzstyle{boundary} = [rectangle, minimum width=0.5cm, minimum height=1cm, text centered, draw=black]
\tikzstyle{arrow} = [thick,->,>=stealth]
\tikzstyle{darrow} = [thick,-,>=stealth]

\node [process,align=center] at (1.5,0) {\Large $\mathcal U$};
\node [process,align=center] at (3.5,0) {\Large$\mathcal U$};
\node  [process,align=center] at (5.5,0) {\Large $\mathcal U$};

\node [process,align=center] at (0.5,1.5) {\Large $\mathcal U$};
\node [process,align=center] at (2.5,1.5) {\Large $\mathcal U$};
\node [process,align=center] at (4.5,1.5) {\Large $\mathcal U$};

\foreach \x in {1,...,6} {
    \draw (\x,-0.5) -- (\x,-0.9);
}

\foreach \x in {0,...,5} {
    \draw (\x,2.0) -- (\x,2.4);
}

\foreach \x in {1,...,5} {
    \draw (\x,0.5) -- (\x,1.0);
}

\node at (0,-1.1) {\large $0$};
\node at (1,-1.1) {\large $1$};
\node at (2,-1.1) {\large $2$};
\node at (5,-1.1) {\large $N$};
\node at (6,-1.1) {\large $N\!+\!1$};
\draw (0,1) --(0,0.4);
\draw[fill=red!40] (0,0.3) circle (0.1);
\node at (0.4,0.3) {\large $\rho_{\rm L}$};
\draw[fill=none] (0,0.0) circle (0.07);
\node at (0.35,-0.1) {\large tr};
\draw (0,-0.07) --(0,-0.9);

\draw (6,2.4) --(6,1.6);
\draw[fill=blue!40] (6,2.4-0.9) circle (0.1);
\node at (6.35,2.4-0.9) {\large $\rho_{\rm R}$};
\draw (6,0.5) --(6,1.08);
\draw[fill=none] (6,1.15) circle (0.07);
\node at (6.35,1.05) {\large tr};

\node at (7,0.75) {\Large $\Rightarrow$};

\node [process,align=center] at (8.5,0) {\Large $\mathcal U$};
\node [process,align=center] at (10.5,0) {\Large$\mathcal U$};
\node [boundary,align=center,fill=blue!40] at (12,0) {\Large $\mathcal{K}_{\rm R}$};

\node [process,align=center] at (9.5,1.5) {\Large $\mathcal U$};
\node [process,align=center] at (11.5,1.5) {\Large $\mathcal U$};
\node [boundary,align=center,fill=red!40] at (8,1.5) {\Large $\mathcal{K}_{\rm L}$};

\foreach \x in {8,...,12} {
    \draw (\x,-0.5) -- (\x,-0.9);
}

\foreach \x in {8,...,12} {
    \draw (\x,2.0) -- (\x,2.4);
}

\foreach \x in {8,...,12} {
    \draw (\x,0.5) -- (\x,1.0);
}

\node at (8,-1.1) {\large $1$};
\node at (9,-1.1) {\large $2$};
\node at (12,-1.1) {\large $N$};

\node at (13.2,0) {\Large $\mathcal{M}_{\rm e}$};
\node at (13.2,1.5) {\Large $\mathcal{M}_{\rm o}$};
\end{tikzpicture}
	\caption{2-step reset driven XYZ circuit in folded notation. Empty circles represent a local traces while red/blue disks represent arbitrary pure boundary states. The right scheme correspond to an equivalent reduced circuit where squares represent boundary Kraus maps (see details in the main text).
}
	\label{fig:scheme}
\end{figure}
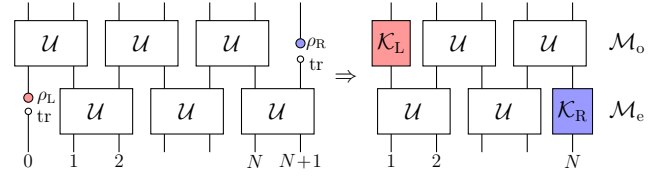

\section{Boundary reset driven XYZ circuits} 

We consider a chain of $N+2$ qubits labelled by $0,1\ldots N,N+1$.  An XYZ gate
is defined as 
\begin{align}
& U(u,\eta)=\left[\tfrac{\bell{1}(\eta)\bell{1}(\eta)}{\bell{1}(\eta+u)\bell{1}(\eta-u)}\right]^{\frac12}\widetilde{U}(u,\eta),
\label{eq:gateU} \\
&\widetilde{U}(u,\eta)=\left(
\begin{array}{cccc}
 r_1 & 0 & 0 & r_4 \\
 0 & r_3 & r_2 & 0 \\
 0 & r_2& r_3 & 0 \\
 r_4 & 0 & 0 & r_1 \\
\end{array}
\right),\\
&r_1= \frac{\el{4}(u)\el{1}(u+\eta)}{\el{4}(0)\el{1}(\eta)},\quad r_2=\frac{\el{1}(u) \el{4}(u+\eta)}{\el{4}(0)\el{1}(\eta)},\\
&r_3= \frac{\el{4}(u)\el{4}(u+\eta)}{\el{4}(0)\el{4}(\eta)},\quad r_4=\frac{\el{1}(u) \el{1}(u+\eta)}{\el{4}(0) \el{4}(\eta)},
\label{Ugate}
\end{align}
where $\el{\al}(u) \equiv \vartheta_\al(\pi u, e^{2 i \pi \tau } )$ and $\bell{\al}(u) \equiv \vartheta_\al(\pi u, e^{ i \pi \tau })$ are standard
elliptic theta-functions, see Appendix \ref{App:theta}.
The $U$ matrix is parametrized by anisotropic parameter $\eta$,  module of the elliptic theta function $\tau$ 
($\imag(\tau)>0$),  and a spectral parameter $u$.  The matrix $P \widetilde{U}$ where $P$ is the permutation operator coincides with the 
standard parametrization of the $R$-matrix for $8$-vertex model \cite{Baxter-book}.  
The operator $U$ satisfies a symmetry relation 
\begin{align} 
&U(u,\eta)= U(-u,-\eta), \label{eq:Usymmetry}
\end{align}
and the unitarity relation 
\begin{align} 
&U(u,\eta) U(-u,\eta)= U(u,\eta) U(u,-\eta)=I.\label{eq:crossing}
\end{align}
For small $u\ll 1$, we have 
\begin{align} 
\widetilde{U}(u,\eta) &= I + u\frac{\bell{1}'(0)}{2 \bell{1}(\eta)}\left(h + \frac{\bell{1}'(\eta)}{\bell{1}'(0)}I \right)+ O(u^2), \label{eq:uToZero}
\end{align}
where $h $ is the energy density of the paradigmatic
XYZ Hamiltonian  
\begin{align}
h_{n,n+1} &= \sum_{\al=1}^3  J_\al \si_n^\al \si_{n+1}^\al.  \label{eq:XYZdensity}
\end{align}
Here $\{\si_n^\al\}$ are Pauli matrices acting on $n$-th spin in the chain,  and the couplings $\{J_\al\}$ read
\begin{align}
 &J_1=\frac{\bell{4}(\eta)}{\bell{4}(0)},\quad J_2=\frac{\bell{3}(\eta)}{\bell{3}(0)},\quad  J_3=\frac{\bell{2}(\eta)}{\bell{2}(0)}.
\label{eq:J}
\end{align}
The couplings $J_\al$ in (\ref{eq:J}) are real if the module of the elliptic theta function $\tau  \in i\mathbb{R}$,  and the anisotropy parameter $\eta \in \mathbb{R}$ or
 $\eta \in i\mathbb{R}$.  These two choices,  after performing the XXZ reduction 
$$\lim_{\tau\to +i\infty}J_{x,y}=1,\quad \lim_{\tau\to +i\infty}J_{z}=\cos(\pi\eta),
$$ correspond to the easy plane $|J_z/J_x|<1$ and easy axis $|J_z/J_x|>1$ regime respectively.  
The $U$ gate  (\ref{eq:gateU}) is unitary  if: (i)
 $u \in \mathbb{R}$,   $\eta \in i\mathbb{R}$  or (ii) $u \in i\mathbb{R}$,   $\eta \in \mathbb{R}$. 


The gate $\cal U$ acts on a pair of neighboring qubits via so-called folded map  as $\mathcal U(\rho) = U \rho U^\dagger$. We consider a chain of $N+2$
qubits labelled as $0,1\ldots N+1$.
Assuming $N$ to be odd (similar construction can be given for $N$ even), 
we define a dynamical protocol consisting of two -- even and odd -- steps,  see Fig.~\ref{fig:scheme}.  During the even step,  we apply a string of  $\cal U$ gates to all neighboring pairs of qubits $(1,2), (3,4),\ldots, (N,N+1)$ while qubit $0$ is traced out and reset to an arbitrary pure state $\rho_{\rm L} = \ket{\psi_{\rm L}}\!\bra{\psi_{\rm L}}$.  During odd step,  we similarly apply $\cal U$ gates to qubits $(0,1),(2,3) \ldots ,(N-1,N)$ and reset the qubit $N+1$ to arbitrary pure state 
$\rho_{\rm R} = \ket{\psi_{\rm R}}\!\bra{\psi_{\rm R}}$.
This defines a  many-body locally interacting (trotterized XYZ) quantum channel,  targeting predefined local spin states at the ends.  Here $u$ can be interpreted as the Trotter time step, so $u\to 0$ corresponds to continuous time limit,  governed by the 
paradigmatic XYZ spin chain Hamiltonian \cite{Baxter-book} (\ref{eq:XYZdensity}) in the bulk,  while  the reset channels 
Fig.~\ref{fig:scheme} corresponds to  Zeno limit of 
instantaneous  relaxation of boundary spins towards targeted states $\rho_{\rm L}$ and $\rho_{\rm R}$.

Ignoring the trivially separable state of the boundary qubit (left/right in even/odd step), we can write an equivalent effective brickwork dynamical quantum channel acting on interior $N$ qubits, labelled $1,2\ldots N$, as
\begin{eqnarray}
\rho_{t+1} &=& \mathcal M( \rho_t),\,\quad\qquad\qquad \mathcal M = \mathcal M_{\rm o} \mathcal M_{\rm e}, \label{eq:M}\\
\mathcal M_{\rm o} &=& \mathcal K_{\rm L} \otimes \mathcal U^{\otimes (N-1)/2}, 
\quad
\mathcal M_{\rm e} = \mathcal U^{\otimes (N-1)/2} \otimes \mathcal K_{\rm R}
, \nonumber
\end{eqnarray}
where ${\mathcal K}_{\rm L/R} : {\rm End}(\mathbb C^2) \to
{\rm End}(\mathbb C^2)$
are the local boundary channels
\begin{align}
\begin{aligned}
\mathcal K_{\rm L}(\rho) &= \tr_1 \left(U (\rho_{\rm L}\otimes \rho)U^\dagger\right),\\
\mathcal K_{\rm R}(\rho) &= \tr_2 \left(U (\rho\otimes \rho_{\rm R})U^\dagger\right)  
\end{aligned}\label{eq:reset}
\end{align}
The boundary channels $\mathcal K_{\rm L/R}(\rho)$  can be expressed explicitly in terms of pairs of Kraus matrices
$\mathcal K_{\rm L/R}(\rho) = \sum_{\mu=1}^2 K^{[\mu]}_{\rm L/R} \rho K^{[\mu]\dagger}_{\rm L/R}$.  
For our further proofs,  an explicit form of Krauss matrices $K^{[\mu]}_{\rm L/R}$ is not needed; for completeness  we 
give them  in the  Appendix~\ref{app:KraussMatrices}.  

\section{Inhomogeneous Yang-Baxter equation}

Let us denote the physical space of $N$ qubits as
$\mathfrak h = (\mathbb C^2)^{\otimes N}$ and label local operators over $\mathfrak h$ which act nontrivially only on $n$-th tensor factor (qubit) by subscript $n$, e.g. Pauli operators $\sigma^{x,y,z}_n$. In addition, we consider an infinite-dimensional auxiliary space $\mathfrak a$ spanned by basis vectors $\{\ket{j}_{\mathfrak a};j=0,1\ldots\}$.
We now define a special type of inhomogeneous Lax operators, and their conjugates, $L^\pm_{\mathfrak a,n},L^{\pm*}_{\mathfrak a,n} \in {\rm End}(\mathfrak a\otimes \mathfrak h)$  as
\begin{eqnarray}
L^+_{\mathfrak a,n}\!\!&=&\!\!\sum_{j=0}^\infty 
\left[\ket{j}\!\bra{j}_{\mathfrak a} 
L_n(n\!-\!2j)
 \!+\! 
\ket{j}\!\bra{j\!+\!1}_{\mathfrak a}
X_n'(n\!-\!2j)\right],\quad\label{eq:Lax} \\
L^-_{\mathfrak a,n}\!\!&=&\!\!\sum_{j=0}^\infty 
\left[\ket{j}\!\bra{j}_{\mathfrak a} 
L_n'(n\!-\!2j)
 + 
\ket{j}\!\bra{j\!+\!1}_{\mathfrak a}
X_n(n-2j)\right], \nonumber \\
L^{+*}_{\mathfrak a,n}\!\!&=&\!\!\sum_{j=0}^\infty
\left[\ket{j}\!\bra{j}_{\mathfrak a} 
L_n(n\!-\!2j)^\dagger 
 + 
\ket{j}\!\bra{j\!+\!1}_{\mathfrak a}
X_n'(n\!-\!2j)^\dagger  \right], \nonumber \\
L^{-*}_{\mathfrak a,n}\!\!&=&\!\!\sum_{j=0}^\infty
\left[\ket{j}\!\bra{j}_{\mathfrak a} 
L_n'(n\!-\!2j)^\dagger
 + 
\ket{j}\!\bra{j\!+\!1}_{\mathfrak a}
X_n(n\!-\!2j)^\dagger \right], \nonumber  
\end{eqnarray}
where $L_n(m),L_n'(m),X_n(m),X_n'(m)$ are rank-$1$ explicit site-dependent matrices, acting on the $n$-th qubit. 

Defining two unnormalized orthogonal qubit states as
\begin{align}
&\ket{\psi(u)} =  \binom{\el{1}(u)}{\el{4}(u)},\quad
\bra{\xi(u)} = (\el{4}(u), \  -\el{1}(u) ), \label{eq:psiXi}
\end{align}
we postulate
\begin{align}
&L(m)=\frac{1}{C(m)} \ket{\psi(m\eta+a)}\bra{\xi(-u-m\eta-a)}, \no\\
 &L'(m)=\frac{1}{C(m)} \ket{\psi(u+m\eta+a)}\bra{\xi(-m\eta- a)},\no \\
&X(m)=\frac{1}{C(m)} \ket{\psi(m\eta-u+a)}\bra{\xi(-m\eta-a)},\label{eq:LaxMatrices}\\
&X'(m)=\frac{1}{C(m)} \ket{\psi(m\eta+a)}\bra{\xi(u-m\eta-a)},\no\\
& C(m)=\bell{2}\left(a+m\eta \right),\no
\end{align}
where $a\in \mathbb C$ is an arbitrary complex parameter.

The following Yang-Baxter-type (or RLL) identities play a key role in the following.
\begin{align}
\begin{aligned}
U_{n,n+1}L^+_{\mathfrak a,n} L^-_{\mathfrak a,n+1} =& 
L^-_{\mathfrak a,n} L^+_{\mathfrak a,n+1} U_{n,n+1}, \\
U_{n,n+1}L^{+*}_{\mathfrak a,n} L^{-*}_{\mathfrak a,n+1} =& 
L^{-*}_{\mathfrak a,n} L^{+*}_{\mathfrak a,n+1} U_{n,n+1},
\end{aligned}
\label{eq:RLL}
\end{align}
where we interpret the gate ${U}_{n,n+1}$ as an R-matrix of a unitary eight-vertex model. 
Contrary to the usual situation,  the RLL relations (\ref{eq:RLL}) are inhomogeneous,  
since our  Lax operators $L^+_{\mathfrak a,n}, \ldots $  (\ref{eq:Lax}) have explicit  $n$-dependence (while the gate $U$ 
remains the same).  A proof of the identities  (\ref{eq:RLL}) is given in the Appendix \ref{App:A}. 



\section{Explicit inhomogeneous matrix product NESS for XYZ circuit}

Similarly to  Ref. \cite{2025XXZcircuit},  we formulate an ansatz for the NESS $\rho_\infty,\rho_\infty'$ appearing at the even and odd steps of 
the qubit circuit evolution
\begin{align}
 &\mathcal M_{\rm o} \mathcal M_{\rm e} \rho_\infty  = \rho_\infty, \label{eq:NESS} \\
&\mathcal M_{\rm e} \mathcal M_{\rm o} \rho_\infty'  = \rho_\infty',   \label{eq:NESSprime}
\end{align}
as follows:
\begin{align}
\rho_\infty &=
 \bra{\rm L}
\mathbb L^+_{\mathfrak{ab},1}\mathbb L^-_{\mathfrak{ab},2} \mathbb L^+_{\mathfrak{ab},3}\cdots
\mathbb L^-_{\mathfrak{ab},N-1}
\mathbb L^+_{\mathfrak{ab},N}
\ket{\rm R},\no\\
\rho'_\infty &=
 \bra{\rm L}
\mathbb L^-_{\mathfrak{ab},1}\mathbb L^+_{\mathfrak{ab},2}\mathbb L^-_{\mathfrak{ab},3}\cdots 
\mathbb L^+_{\mathfrak{ab},N-1}
\mathbb L^-_{\mathfrak{ab},N}
\ket{\rm R},
 \label{eq:MPA}\\
\mathbb L^\pm_{{\mathfrak a}{\mathfrak b},n} &= 
L^\pm_{\mathfrak a,n}
L^{\pm*}_{\mathfrak b,n}, \no
\end{align}
where $\bra{\rm L}$, $ \ket{\rm R}$ are auxiliary vectors in a product of two formally unbounded auxiliary spaces ${\mathfrak a} \otimes{\mathfrak b}$, i.e.  $\bra{\rm L}=\sum_{j,j'=0}^\infty l_{j,j'} \bra{j}_{\mathfrak{a}}\bra{j'}_{\mathfrak{b}}$,  and 
$\ket{\rm R}=\sum_{j,j'=0}^\infty r_{j,j'} \ket{j}_{\mathfrak{a}}\ket{j'}_{\mathfrak{b}}$.  
However,  our  solution for $\bra{\rm L}$ has just few nonzero components $l_{jj'}$, see (\ref{eq:LEasyPlane}), (\ref{eq:LEasyAxis}), 
and therefore, generic (\ref{eq:MPA}) always contains finite number of 
terms, see more details after Eq.(\ref{eq:nu}).

Consistency of the ansatz (\ref{eq:MPA}) in the bulk is guaranteed by Eq. (\ref{eq:RLL}),  while the boundaries yield extra conditions to determine 
$\bra{\rm L}$ and $\ket{ \rm R}$:
\begin{eqnarray}
&&\bra{\rm L}\left(\mathbb L^+_{\mathfrak{ab},1}-
\mathcal{K}_{\rm L}(\mathbb L^{-}_{\mathfrak{ab},1})\right) = 0\,, \label{eq:LBE}\\
&&
\left(\mathbb L^-_{\mathfrak{ab},N}-
\mathcal{K}_{\rm R}(\mathbb L^+_{\mathfrak{ab},N})\right)\ket{\rm R} = 0\,.
\label{eq:RBE}
\end{eqnarray}

Like (\ref{eq:RLL}),  the conditions (\ref{eq:LBE}) and (\ref{eq:RBE}) are overdetermined.  Solutions of these equations
depends on the reset channels (\ref{eq:reset}) which will be 
 parametrized  with two complex numbers 
$\al_{\rm L}$,  and $\al_{\rm R}$, 
\begin{align}
\begin{aligned}
&\rho_{\rm L} = \ket{\psi_{\rm L}}\bra{\psi_{\rm L}}, \quad \rho_{\rm R} = \ket{\psi_{\rm R}}\bra{\psi_{\rm R}}. \\
&\ket{\psi_{\rm L}}=\frac{1}{\sqrt{\bell{4}(\real(\al_{\rm L}))\bell{3}(i\real(\al_{\rm L}))}}\binom{\th_1(\al_{\rm L})}{\th_4(\al_{\rm L})},\\
&\ket{\psi_{\rm R}}=\frac{1}{\sqrt{\bell{4}(\real(\al_{\rm R}))\bell{3}(i\real(\al_{\rm R}))}}\binom{\th_1(\al_{\rm R})}{\th_4(\al_{\rm R})},
\end{aligned}s
\label{eq:psiLR}
\end{align}
The $U$ gate  (\ref{eq:gateU}) is unitary  if either
 $u \in \mathbb{R}$,   $\eta \in i\mathbb{R}$  or  $u \in i\mathbb{R}$,   $\eta \in \mathbb{R}$.  The two cases 
are treated separately below. The  proofs of following Eqs. (\ref{eq:AuxiliaryVecL})-(\ref{eq:CnmCaseA}) are given in 
Appendices \ref{App:left} and \ref{App:right}.

\textit{Case    $\eta \in \mathbb{R}$,  $u \in i\mathbb{R}$.~~ }
The NESS of the XYZ qubit circuit in Fig. \ref{fig:scheme} is given by Eq. (\ref{eq:MPA})
with local entries of Lax matrix given by (\ref{eq:LaxMatrices}), where $a\equiv \al_{\rm L}=w_0+ i w_1$,
and the auxiliary vector $\bra{\rm L}$ has \textit{just $4$ nonzero components},
\begin{align}
\bra{\rm L}&=\sum_{j,j'=0}^1l_{j,j'} \bra{j}_{\mathfrak{a}}\bra{j'}_{\mathfrak{b}},  \label{eq:AuxiliaryVecL}\\
l_{0,0}&=1,\quad
l_{1,1}=\frac{\bell{3}(u-i w_1) }{\bell{3}(u+i w_1)}, \label{eq:LEasyPlane}\\
l_{0,1}&=\frac{\bell{3}(i w_1) \bell{4}(u+ w_0) }{\bell{3}(u+i w_1)\bell{4}( w_0) }, \quad l_{1,0}=l_{0,1}^*, \no\\
w_0&={\rm Re}[\al_{\rm L}],\quad w_1={\rm Im}[\al_{\rm L}].\no
\end{align}
The auxiliary vector $\ket{\rm R}$ is unbounded,  and has components 
\begin{align}
&\ket{\rm R}=\sum_{j,j'=0}^\infty r_{j,j'} \ket{j}_{\mathfrak{a}}\ket{j'}_{\mathfrak{b}}, 
\label{eq:AuxiliaryVecR}\\
&r_{j,j'} = C_{j,j'} \prod_{k=0}^{j-1} b_k  \prod_{k'=0}^{j'-1} b_{k'}^*, \label{RAnsatz}\\
& C_{j,j'} = \frac{\bell{3}((j-j')\eta-i \Ga_1) \,
\bell{4}((j+j')\eta-   \Ga_0) }{ \bell{3}(i \Ga_1) \, \bell{4}(\Ga_0) },\label{eq:CnmCaseB} \\
&b_n=  -\frac{\bell{1}(n\eta +\nu) \, \bell{2}(u+\Ga- n\eta+\nu) }
{\bell{1}(u+(n+1)\eta+\nu)\,  \bell{2}(\Ga-(n+1)\eta+\nu) }, 
\label{eq:bn}\\
&\Ga =  \al_{\rm L}+(N+1)\eta,\quad \Ga_0={\rm Re}[\Ga], \ \Ga_1={\rm Im}[\Ga], \\
&\nu= \frac12 \left( \al_{\rm R} -\al_{\rm L}- u-  (N+1)\eta \right).\label{eq:nu}
\end{align}
Even though the vector $\ket{\rm R}$  is formally unbounded,  for a  system of size $N$,   only the finite subset of coefficients  $r_{j,j'}$
with $j,j' \leq N+2$ matters.  Indeed,  due to the band structure of Lax operators, 
the expression in (\ref{eq:MPA}) 
$\bra{\rm L}
\mathbb L^+_{\mathfrak{ab},1}\mathbb L^+_{\mathfrak{ab},2} \ldots $  can be seen as a walk on 
$2$-dimensional graph with coordinates $j,j'\geq 0$   starting from a $2\times 2$ seed (\ref{eq:AuxiliaryVecL}) at the origin.   Every application of
$\mathbb L^+_{\mathfrak{ab},n}$  increases the walk range   by at most  one unit $j\rightarrow j, j+1$,  
$j'\rightarrow j',j'+1$.  After $N$ steps,  the walk can only reach the points $j,j'\leq N+2$ on the graph,    making the elements of 
$r_{j,j'}$ outside the walk range   obsolete.

\textit{Case    $\eta \in i\mathbb{R}$,  $u \in \mathbb{R}$.~~ }
The NESS of the XYZ qubit circuit in Fig. \ref{fig:scheme} is given by Eq. (\ref{eq:MPA})
with  Lax matrices (\ref{eq:LaxMatrices}),  where $a\equiv \al_{\rm L}=w_0+ i w_1$,
and the auxiliary vector $\bra{\rm L}$  (\ref{eq:AuxiliaryVecL})  with
\begin{align}
&l_{0,0}=1, \quad l_{1,1}=    \frac{\bell{4}(u-w_0) }{\bell{4}(u+ w_0) },\no\\
&l_{0,1}=  \frac{\bell{3}(u+ i w_1) \bell{4}( w_0) }{\bell{3}(i w_1)\bell{4}(u+ w_0) }, \quad l_{1,0}=l_{0,1}^*,
\label{eq:LEasyAxis}
\end{align}
while the right boundary vector $\ket{\rm R}$ has the form (\ref{eq:AuxiliaryVecR}),  (\ref{RAnsatz}) 
with $b_n$
given by the  expression (\ref{eq:bn}) and  $C_{j,j'}$ given by 
\begin{align}
& C_{j,j'} = \frac{\bell{4}((j-j')\eta-\Ga_0) 
\bell{3}((j+j')\eta- i  \Ga_1) }{\bell{4}(\Ga_0) \bell{3}(i \Ga_1) }.\label{eq:CnmCaseA} 
\end{align}

Notably,  both left and right auxiliary vectors  generally  cannot be recast in a product form,
$\bra{\rm L}_{\mathfrak{a b}} \neq \bra{v}_{\mathfrak{a}}\bra{\tilde{v}}_{\mathfrak{b}}$,
$\ket{\rm R}_{\mathfrak{a b}} \neq \ket{w}_{\mathfrak{a}}\ket{\tilde{w}}_{\mathfrak{b}}$.  Consequently,  the NESS of a driven qubit circuit 
(\ref{eq:MPA}) cannot be factorized into a product of two Cholesky-type 
factors as it was in all previous solvalble instances in models with continuous  time evolution ~\cite{TopicalReview}.  
Apparently,  in absence of $U(1)$ symmetry,  it is the time discreteness,  i.e.  a finite Trotter time step $u \neq 0$ in (\ref{eq:gateU})
which makes the coupling between auxiliary spaces $\mathfrak{a},\mathfrak{b}$ a necessary ingredient of a theory.  
In special cases factorization can still happen,  e.g.  for $u=0$ or for $u_{\rm L} = \frac12$ the left auxiliary vector 
$\bra{\rm L}$ is factorized,  while a 
generic infinite-dimensional $\ket{{\rm R}}$  
is factorized only in a trivial case $\eta=0$.  Generic Schmidt rank for both auxiliary vectors is equal to $2$. 
However,  for   specially tuned boundary conditions the vector $\ket{{\rm R}}$ has a finite support leading to special 
class of robust circuit states -- brickwork helices and their descendants -- which are described below.   

\section{Elliptic spin helices in XYZ brickwork circuits}

An analysis of  Eqs.(\ref{eq:AuxiliaryVecR}), (\ref{eq:bn})  reveals an especially simple class of brickwork NESS  
(\ref{eq:NESS}): it corresponds to   zeros and poles of $b_n$.  Namely,  setting $\nu=0$,
or 
\begin{align}
&\al_{\rm R}=\al_{\rm L}+(N+1)\eta +u,   \label{eq:helixCondition}
\end{align}
leads to $b_0=0$ and consequently 
$\ket{\rm R} =  \ket{0}_{\mathfrak{a}}\ket{0}_{\mathfrak{b}}$.
 Due to upper triagonal  structure of the Lax matrices (\ref{eq:Lax}), 
the NESS (\ref{eq:MPA}) then simplifies to  
\begin{align}
\rho_\infty &= L(1)L(1)^\dagger \otimes  L'(2)L'(2)^\dagger \otimes \ldots \otimes  L(N)L(N)^\dagger\no\\
&= A \ket{\Psi_{+}} \bra{\Psi_{+}}, \no\\
\ket{\Psi_{+}} &=\psi_{1}(0) \otimes \psi_{2}(u) \otimes \psi_{3}(0)\otimes \psi_{4}(u)\no\\
&\quad\dots \otimes  \psi_{N-1}(u) \otimes  \psi_{N}(0),\label{eq:SHSplus}\\
\rho_\infty' &= A \ket{\Psi_{+}'} \bra{\Psi_{+}'},\no\\
\ket{\Psi_{+}'} &=\psi_{1}(u) \otimes \psi_{2}(0) \otimes \psi_{3}(u)\otimes \psi_4(0)\no\\
&\quad\dots \otimes  \psi_{N-1}(0)\otimes  \psi_{N}(u),\label{eq:SHSplusPrime}
\end{align}
where  $A$ is a normalization constant and 
\begin{align}
&\psi_n(u) = \binom {\th_1(\al_{\rm L}+n \eta +u)} {\th_4(\al_{\rm L}+n \eta+u )}. \label{eq:psi(u)}
\end{align}

\begin{figure}[tbp]
  \centering
  \includegraphics[width=0.45\textwidth]{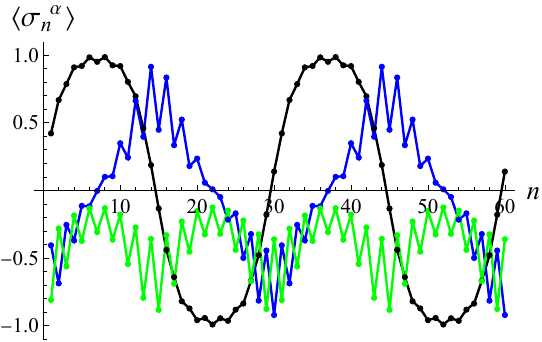}
 \includegraphics[width=0.45\textwidth]{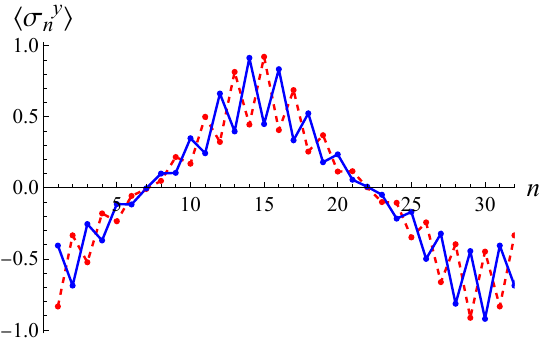}
 \caption{
Average local magnetizations profiles for elliptic brickwork helix  (\ref{eq:SHSplusPeriodic})
for real $\eta$.
Top panel: black, blue, green points connected with full lines correspond to 
$\langle \si_n^x \rangle$, $\langle \si_n^y\rangle$, $\langle \si_n^z\rangle$  respectively.      
Bottom panel:  Blowup of  $\langle \si_n^y\rangle$ for even/odd steps of evolution 
(\ref{eq:EvenEvolutionStep}), (\ref{eq:OddEvolutionStep}) is 
 shown with full blue and dashed red lines respectively. 
Parameters: $M=60,  u =0.15 i, \tau=0.65 i,  \al_{\rm L}= 0.095 +  0.1 i$,  $\eta = 4/M$. 
   }
  \label{FigSHSellipticEasyPlane}
\end{figure}

\begin{figure}[tbp]
  \centering
  \includegraphics[width=0.45\textwidth]{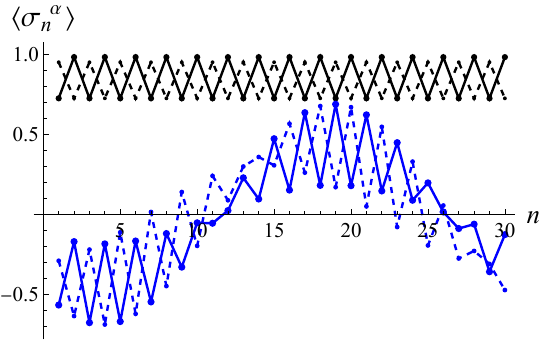}
 \caption{
Average local magnetizations profiles for elliptic brickwork helix  (\ref{eq:SHSplusPeriodic})
for  imaginary $\eta$.  Black and blue points connected with full lines correspond to 
$\langle \si_n^x \rangle$ and $\langle \si_n^y\rangle$, respectively.      
Dashed lines show the same quantity at odd steps of evolution (\ref{eq:OddEvolutionStep}).  
Parameters: $M=30,  u =0.15, \tau=0.325 i,  \al_{\rm L}= 0.095 +  0.1 i$,  $\eta =2 \tau/M$. 
   }
  \label{FigSHSellipticEasyAxis}
\end{figure}

The state (\ref{eq:SHSplus}) is periodic in space (with lattice period $2/\eta$ or $2\tau/\eta$,  depending on whether the anisotropy $\eta$ is real or imaginary), 
with qubit polarization vector following  an elliptic helicoidal spiral. The argument of elliptic function grows linearly by unit $\eta$ with spin position,
  with an additional even/odd site  staggering due to additional argument increment ``$+u$" for even sites. 
Typical local magnetization profiles for elliptic helices are shown in Figs~\ref{FigSHSellipticEasyPlane}, \ref{FigSHSellipticEasyAxis}.
For  odd steps of evolution,  the NESS  is still fully factorized, as in Eq. (\ref{eq:SHSplusPrime}), but the increment ``$+u$" occurs at odd sites,  which 
results in different  magnetization profiles at odd and even steps of qubit circuit.

Significance of the factorized helix state (\ref{eq:SHSplus}) goes beyond just being a special fine-tuned solution of the steady state equation 
(\ref{eq:NESS}).  
A particularity of  helix (\ref{eq:SHSplus}) is that it is locally 
reproduced  in the bulk after every full step $ \mathcal M_{\rm o} \mathcal M_{\rm e}$.  This
renders it particularly robust also  in open circuits of size $N$,  with arbitrary boundary conditions, 
where spin helix lifetime would be of order  $N$,
while a lifetime of a generic state is of order $O(1)$.  In the XXZ limit  $\imag (\tau) \to \infty$ and for real $\eta$,  the state (\ref{eq:SHSplus}) 
describes winding of the polarization vector  along the chain in harmonic fashion,  with even-odd site alternation, see \cite{2025XXZcircuit}.   The  usefulness of the  long-lived helices  for calibration purposes has been 
demonstrated in cold atoms experiments  ~\cite{KettNature2022}.  
Further we shall refer to the state (\ref{eq:SHSplus}) as an elliptic (XYZ) brickwork spin helix,  or simply a brickwork helix.
Note that  periodic in space helix (\ref{eq:SHSplus})
with imaginary $\eta$ and lattice period $2\tau/ \eta$ does not have a trigonometric XXZ analog.  
Brickwork helix state also appears as a special degenerate site-factorized stationary state for periodic circuits  (see (\ref{eq:SHSplusPeriodic}),   (\ref{eq:SHSminusPeriodic})).

Notably,  slight generalization of (\ref{eq:helixCondition})
 \begin{align}
&\al_{\rm R}=\al_{\rm L} + (N+1-2 m_{\rm k})\eta  +u,   \label{eq:KinkCondition}\\
&m_{\rm k}=0,1,2,\ldots \no
\end{align} 
with integer $m_{\rm k}=0,1,2,\ldots$ leads to $b_{m_{\rm k}} =0$ and 
cropping of $\ket{\rm R}$ to dimension $(m_{\rm k}+1)\times (m_{\rm k}+1)$, i.e. 
$r_{j,j'}=0$ if $j>m_{\rm k}$ or $j'>m_{\rm k}$.
The respective NESS appears to be no longer pure (like (\ref{eq:SHSplus})) but it still has a 
low rank,  and it can be seen as consisting from  imperfect helices with $m_{\rm k}$ kinks,  in analogy to~\cite{PopkovSHSkinks}.  The NESS ranks of $\rho_\infty$ for the cases $m_{\rm k}=0,1,2$ cases are 1, $N+1$ and $\binom{N+1}{2}+1$, respectively.  

An elliptic brickwork helix NESS with a winding in an opposite direction
\begin{align}
\ket{\Psi_{-}} &=\psi_{-1}(0) \otimes \psi_{-2}(-u) \otimes \psi_{-3}(0)\otimes \psi_{-4}(-u)\no\\
&\quad \dots \otimes  \psi_{-N+1}(-u) \otimes  \psi_{-N}(0), \label{eq:SHSminus}
\end{align}
appears for a choice of parameters 
 \begin{align}
&\al_{\rm R}=\al_{\rm L} -(N+1)\eta  -u,   \label{eq:helixMinusCondition}
\end{align}
(compare to (\ref{eq:helixCondition})),  and it corresponds to a pole   $1/b_N =0$ in  (\ref{eq:bn}). 
Then,  a  renormalized right auxiliary 
vector $\ket{\rm R}$  has  just one nonzero component,   $\ket{\rm R} =  \ket{N+1}_{\mathfrak{a}}\ket{N+1}_{\mathfrak{b}}$.
The   NESS (\ref{eq:MPA}) then becomes
\begin{align}
\rho_\infty &=l_{11}  \bra{1}_{\mathfrak{a}}\bra{1}_{\mathfrak{b}}      \mathbb L^+_{\mathfrak{ab},1}\mathbb L^-_{\mathfrak{ab},2}
\ldots \ket{N+1}_{\mathfrak{a}}\ket{N+1}_{\mathfrak{b}}
\no\\
&=l_{11} X'(-1)X'(-1)^\dagger \otimes  X(-2)X(-2)^\dagger \otimes \no \\
&\quad \dots \otimes X(1-N)X(1-N)^\dagger \otimes   X'(-N)X'(-N)^\dagger\no\\
&= A \ket{\Psi_{-}} \bra{\Psi_{-}},
\end{align}
where  $A$ is a normalization constant and $\ket{\Psi_{-}}$ is given by (\ref{eq:SHSminus}).
A generalized condition  
 \begin{align}
&\al_{\rm R}=\al_{\rm L} - (N+1-2 m_{\rm k})\eta  -u ,  \label{eq:KinkMinusCondition}
\end{align} 
(compare to (\ref{eq:KinkCondition})) corresponds to a pole $1/b_{N-m_{\rm k }} =0$ and yields
NESS in the form of imperfect helix with $m_{\rm k}$ kinks.  

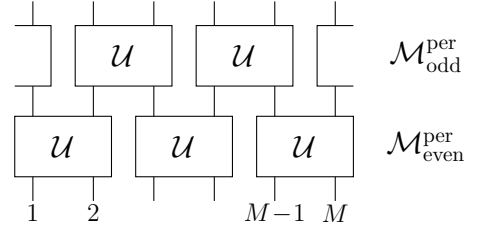
\begin{figure}[htbp]
\centering
\begin{tikzpicture}[node distance=2cm, scale=0.8, transform shape]
\tikzstyle{process} = [rectangle, minimum width=1.6cm, minimum height=1cm, text centered, draw=black]
\tikzstyle{boundary} = [rectangle, minimum width=0.5cm, minimum height=1cm, text centered, draw=black]
\tikzstyle{arrow} = [thick,->,>=stealth]
\tikzstyle{darrow} = [thick,-,>=stealth]

\node [process,align=center] at (1.5,0) {\Large $\mathcal U$};
\node [process,align=center] at (3.5,0) {\Large$\mathcal U$};
\node  [process,align=center] at (5.5,0) {\Large $\mathcal U$};

\node [process,align=center] at (2.5,1.5) {\Large $\mathcal U$};
\node [process,align=center] at (4.5,1.5) {\Large $\mathcal U$};

\foreach \x in {1,...,6} {
    \draw (\x,-0.5) -- (\x,-0.9);
}

\foreach \x in {1,...,6} {
    \draw (\x,2.0) -- (\x,2.4);
}

\draw (1.5-0.8,2)--(1.3,2)--(1.3,1)--(0.7,1);
\draw (5.5+0.8,2)--(5.7,2)--(5.7,1)--(6.3,1);

\foreach \x in {1,...,6} {
    \draw (\x,0.5) -- (\x,1.0);
}

\node at (1,-1.1) {\large $1$};
\node at (2,-1.1) {\large $2$};
\node at (5,-1.1) {\large $M\!-\!1$};
\node at (6,-1.1) {\large $M$};

\node at (7.5,0) {\Large $\mathcal{M}^{\rm per}_{\rm even}$};
\node at (7.5,1.5) {\Large $\mathcal{M}^{\rm per}_{\rm odd}$};

\end{tikzpicture}
	\caption{Two-step reset driven periodic XYZ circuit in folded notation.
}
	\label{fig:scheme:periodic}
\end{figure}

\emph{Divergence condition---}
Bulk stationarity of  elliptic brickwork helices (\ref{eq:SHSplus}),  (\ref{eq:SHSminus}) also follows in a simple way, 
via two Yang-Baxter RLL-type relations,
\begin{align}
&U \ket{\psi_n(0)}\bra{\psi_n(0)}\otimes  \ket{\psi_{n+1}(u)}\bra{\psi_{n+1}(u)}  \no\\
&=\ket{\psi_n(u)}\bra{\psi_n(u)}\otimes  \ket{\psi_{n+1}(0)}\bra{\psi_{n+1}(0)}U,
\label{eq:RLL1}\\
&U \ket{\psi_n(0)}\bra{\psi_n(0)}\otimes  \ket{\psi_{n-1}(-u)}\bra{\psi_{n-1}( -u)}   \no\\
&=\ket{\psi_n(-u)}\bra{\psi_n(-u)}\otimes  \ket{\psi_{n-1}(0)}\bra{\psi_{n-1}(0)}U.
\label{eq:RLL2}
\end{align}
which can be derived from Eq. \eqref{App:A1}.  
The  RLL-type relations (\ref{eq:RLL1}), (\ref{eq:RLL2}) guarantee the stationarity of XYZ brickwork circuits even in absence of  any dissipative Krauss maps.  
Consider  a space-periodic circuit of even size $M$,  $M+1 \equiv 1$ with fully coherent discrete time evolution dynamics
\begin{align}
&\rho_{t+1}= \mathcal M^{\rm per} (\rho_t)=  \mathcal M^{\rm per}_{\rm odd} \mathcal M^{\rm per}_{\rm even} \rho_t, \label{eq:Mperiodic}\\
&\mathcal M^{\rm per}_{\rm odd}  = \mathcal {U}_{M,1}  \mathcal {U}_{2,3} \ldots \mathcal {U}_{M-2,M-1},\\
&\mathcal M^{\rm per}_{\rm even}= \mathcal {U}_{1,2}  \mathcal {U}_{3,4} \ldots \mathcal {U}_{M-1,M},
\end{align}
see Fig. ~\ref{fig:scheme:periodic}. 

If a periodic closure condition is satisfied
\begin{align}
&\eta M = 0 \quad \mbox{mod}\,\,2, \quad \mbox{if $\eta \in \mathbb{R}$},\\
 &\eta M = 0 \quad \mbox{mod} \,\, 2\tau, \quad \mbox{if $\eta \in i\mathbb{R}$},
\end{align}
then  an even-size version of Eqs.~(\ref{eq:SHSplus}),~(\ref{eq:SHSminus})
\begin{align}
\ket{\Psi_{+}} &=\psi_{1}(0) \otimes \psi_{2}(u) \otimes \psi_{3}(0)\otimes \otimes \psi_{4}(u)\no\\
&\quad \dots \otimes  \psi_{M-1}(0) \otimes  \psi_{M}(u),\label{eq:SHSplusPeriodic}\\
\ket{\Psi_{-}}&=\psi_{-1}(0) \otimes \psi_{-2}(-u) \otimes \psi_{-3}(0)\otimes \otimes \psi_{-4}(-u)\no\\
&\quad \dots \otimes  \psi_{-M+1}(0) \otimes  \psi_{-M}(-u), \label{eq:SHSminusPeriodic}
\end{align}
satisfy stationarity conditions 
\begin{align}
&\mathcal M^{\rm per}_{\rm odd} M^{\rm per}_{\rm even} (\ket{\Psi_{\pm}}  \bra{\Psi_{\pm}}) = \ket{\Psi_{\pm}}  \bra{\Psi_{\pm}}, \label{eq:EvenEvolutionStep}\\
&\mathcal M^{\rm per}_{\rm even} \mathcal M^{\rm per}_{\rm odd} (\ket{\widetilde{\Psi}_{\pm}}  \bra{\widetilde{\Psi}_{\pm}}) = 
\ket{\widetilde{\Psi}_{\pm}}  \bra{\widetilde{\Psi}_{\pm}},\no\\
&\ket{\widetilde{\Psi}_{\pm}}  \bra{\widetilde{\Psi}_{\pm}} =  M^{\rm per}_{\rm even}  \ket{\Psi_{\pm}}  \bra{\Psi_{\pm}}
\label{eq:OddEvolutionStep}
\end{align}
where $\ket{\widetilde{\Psi}_{\pm}}$ are obtained from $\ket{\Psi_{\pm}}$ by swapping the arguments $0\leftrightarrow \pm u$
in  (\ref{eq:SHSplusPeriodic}), (\ref{eq:SHSminusPeriodic}).  Equations (\ref{eq:EvenEvolutionStep})-(\ref{eq:OddEvolutionStep})
follow
straightforwardly from   (\ref{eq:RLL1}), (\ref{eq:RLL2}). 
Stationary helices  (\ref{eq:SHSplusPeriodic}), (\ref{eq:SHSminusPeriodic}) in the 
periodic setup (\ref{eq:Mperiodic}) are degenerate, since the choice of the overall phase $\al_{\rm L}$ in  (\ref{eq:psi(u)}) is arbitrary.  
On a base of  numerical investigations, we propose that the degeneracy of a helices 
in a periodic system of $M$ sites (see Fig. ~\ref{fig:scheme:periodic}), 
  including both chirality signs, 
is equal $\deg= 2M$. 
Helices  (\ref{eq:SHSplusPeriodic}), (\ref{eq:SHSminusPeriodic}) for $\eta \in i \mathbb{R}$ do not have analogs in XXZ-type brickwork circuits \cite{PopkovSHSkinks}.
Figs.~\ref{FigSHSellipticEasyPlane} and \ref{FigSHSellipticEasyAxis} show examples of magnetization profiles for the periodic 
elliptic helices (\ref{eq:SHSplusPeriodic}) for both  $\eta \in  \mathbb{R}$ and  $\eta \in i \mathbb{R}$ cases.

\emph{Visualization of elliptic spin helices via  measuring one-point correlations---}
Coming back to the open circuit with boundary resets, Fig.~\ref{fig:scheme}, we propose to observe 
the elliptic brickwork helices in a potential experiment,  where we change the anisotropy and 
measure a one-point correlation at a  boundary.    

To this end,  let us 
fix the left/right boundary resets as in  (\ref{eq:psiLR}) with arbitrary $\al_{\rm L}$,  and  $\al_{\rm R}=\al_{\rm L} +u$.
Keeping the resets fixed,  we adiabatically change the anisotropy $\eta$.  
The condition (\ref{eq:helixCondition}) predicts pure helix NESS for the  anisotropy satisfying 
 \begin{align}
& (N+1)\eta  =2n\tau, \quad n=1,2,\ldots ,N,\quad \mbox{for case (A)}, \label{eq:helixCondA}\\
& (N+1)\eta  = 2n,  \quad n=1,2,\ldots ,N, \quad  \mbox{for case (B)}. \label{eq:helixCondB}
\end{align}

For experimental detection of  helices,  we 
propose to measure the one-point steady state correlations 
of the leftmost spin: $a=\langle \sigma_1^+ \rangle$ and 
$b=\langle \sigma_1^z \rangle$,  from which the reduced density matrix of the leftmost spin
$\rho_1 = {\rm tr}_{2,3,\ldots, N} (\rho_{\infty})$ can be constructed.
Next, we define 
 two helix  indicators: 
\begin{align}
&f_{1} = 1-{\rm tr} (\rho_1^2) = \frac12 - \frac{b^2}{2} -2 a a^*, \label{def:f1}\\
&f_{2}(x)= \frac12 {\rm tr} |\rho_1 - \rho(\al_{\rm L}+x)| = \frac12 \sum_{j=1}^2 |\la_j|^2,
\label{def:f2}\\
&\rho(x) =\frac{1}{ |\el{1}(x)|^2 + |\el{4}(x)|^2} \binom{\el{1}(x)}{\el{4}(x)} \left( \el{1}(x^*),\,\el{4}(x^*) \right).
\label{eq:targetSpin1}
\end{align}
The indicator $f_{1}$ simply measures a purity of a qubit state.  
The indicator $f_2(\eta)$   is a trace distance between the actual state of the first qubit $\rho_1$ and $\rho(\al_{\rm L}+\eta)$,
which is  the  state of the first qubit in the helix (\ref{eq:SHSplus}).
Analogously, $f_2(-\eta)$  measures a distance between $\rho_1$ and  respective state $\rho(\al_{\rm L}-\eta)$ in the helix of opposite chirality (\ref{eq:SHSminus}).
In Eq. (\ref{def:f2}),  $|\la_j|$  are absolute values of the eigenvalues of the difference $\rho_1-\rho(\al_{\rm L}+x)$.
By definition,  $f_1,f_2$ have range $0\leq  f_1\leq  1/2$,  and  $0\leq  f_2\leq  1$.
Whenever the NESS becomes an elliptic helix state (\ref{eq:SHSplus}),  both indicators vanish,   $f_1=f_2(\eta)=0$.
This is expected to happen whenever the conditions  (\ref{eq:helixCondA}) or (\ref{eq:helixCondB})
are satisfied. Otherwise  $f_{1},f_{2}(\pm \eta)$ serve as a measure of a purity and of a distance to an elliptic helices
(\ref{eq:SHSplus}), (\ref{eq:SHSminus}) respectively.

\begin{figure}[htbp]
  \centering
  \includegraphics[width=0.45\textwidth]{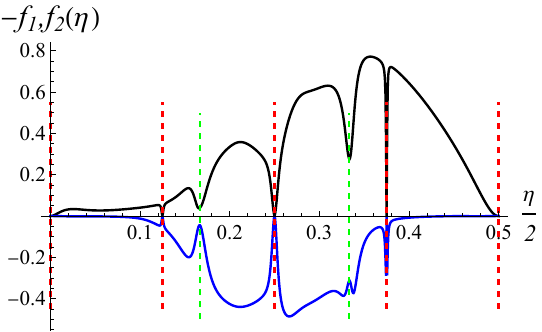}
 \includegraphics[width=0.45\textwidth]{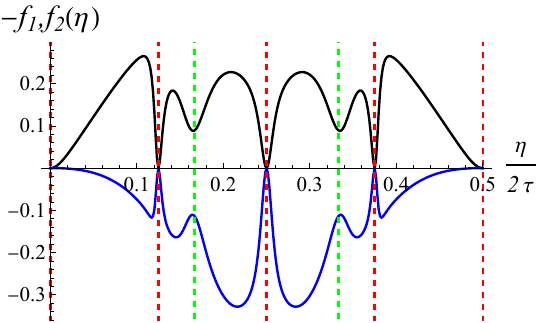}
 \caption{Helix indicators $-f_1$ (solid blue curve) and $f_2(\eta)$ (solid black curve), given by (\ref{def:f1}) and (\ref{def:f2}) respectively and measured in the NESS, plotted versus the rescaled anisotropy for a system of $N = 7$ sites.
Top Panel: $u =0.185 i, \tau=0.65 i,  \al_{\rm L}= 0.165 +  0.13 i, \, \al_{\rm R} =  \al_{\rm L} +u$.  
Bottom Panel: $u =0.185, \tau=0.65 i,  \al_{\rm L}= 0,\al_{\rm R} = u=0.185$. 
Zeros of the indicators coincide with the pure helix condition 
$(N+1)\eta = 0\pmod{2}$, 
(red dashed lines),  while  less pronounced minima (green dashed  lines) occur 
at  anisotropies   leading to  helix with integer number of kinks  $m_{\rm k}=1$ (helix descendants)
$(N+1-2 m_{\rm k})\eta = 0$,  $m_{\rm k}=1$.
   }
  \label{FigPurityBA7}
\end{figure}

\begin{figure}[htbp]
  \centering
  \includegraphics[width=0.45\textwidth]{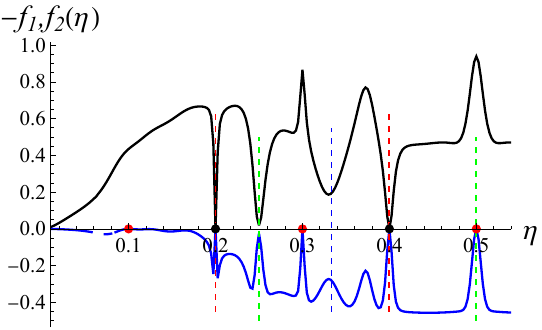}
 \includegraphics[width=0.45\textwidth]{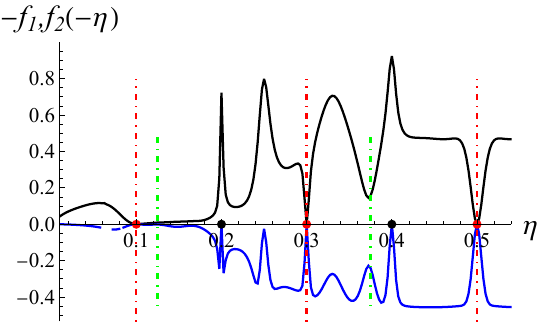}
 \caption{Helix indicators $-f_1$ (blue curve),  $f_2(\pm \eta) $ (black curve) given by (\ref{def:f1}), (\ref{def:f2}), 
versus the  anisotropy,  for system with $N=9$ sites.  Zeros of $f_1$ are marked by black and red points. 
Parameters:  $u =0.185 i, \tau=0.65 i,  \al_{\rm L}=0, \, \al_{\rm R}=\al_{\rm L}+u$.  
Top Panel: a  subset of zeros of $f_1$ (black points ) coincides with the pure helix condition 
$(N+1)\eta = 0\pmod{2 }$, and corresponds to helices of type (\ref{eq:SHSplus})
(red dashed lines).  Helix descendants with  one/two kinks correspond to 
$(N+1-2 m_{\rm k})\eta = 0\pmod{2}$ with $m_{\rm k}=1,2$,  and are marked with green/blue dashed  lines.  
Red points mark location of  helices with opposite helicity of the form (\ref{eq:SHSminus}),
appearing at $(N+1)\eta = 1\pmod{2 }$,  see text for more explanations.
Bottom Panel: remaining zeros of $f_1$ (red points) correspond to helices of reversed chirality (\ref{eq:SHSminus}) and their descendants
and appear at positions $(N+1-2 m_{\rm k})\eta = 1\pmod{2 }$ with $m_{\rm k}=0,1$,  marked with 
red and green dotdashed lines.
   }
  \label{FigPurityB09}
\end{figure}

Figure \ref{FigPurityBA7} shows  that  indicators $f_{1},f_2(\eta)$
are very sensitive not only to the presence of the helices (at positions (\ref{eq:helixCondA}), (\ref{eq:helixCondB})
as expected),  but they also show very clear
 less deep minima (indicated by green dashed lines). 
Positions of the extra minima can be readily identified as 
location of  helix descendants,  i.e., helices with kinks,  according to (\ref{eq:KinkCondition}).
For the case $N=7$, we see clear minima of $f_1,f_2$  for helix descendants  with $m_{\rm k}=1$.  
For an increasing system size,   more and more helix descendants $m_{\rm k}\geq 1$ can be identified,  see 
Fig.~\ref{FigPurityB09}.

In Fig.~\ref{FigPurityBA7} only helices with the same  winding direction   appear,  the helices  of type  (\ref{eq:SHSplus}).  
However,  with a specific choice of boundary condition $\al_{\rm L},  \, \al_{\rm R}$, and only in 
 XYZ circuits,   by varying anisotropy one can  generate  NESS with any winding direction,  i.e. 
both of (\ref{eq:SHSplus})-type and of (\ref{eq:SHSminus})-type.  This becomes possible due to a specific property of 
theta functions $\th_1,\th_4$ parametrizing our  reset channels,  namely 
\begin{align}
& \binom{\th_1(\al)}{\th_4(\al)} = \binom{\th_1(1-\al)}{\th_4(1-\al)}. 
\label{eq:property1/2}
\end{align}

A scenario featuring steady helices of both chiralities is illustrated in Fig. \ref{FigPurityB09}.  
Namely,  choosing  the left and right reset as 
$\al_{\rm L} =0, \al_{\rm R} =u$,  we generate NESS -  helices of type (\ref{eq:SHSplus}) when the condition (\ref{eq:helixCondition}), 
equivalent to  $(N+1)\eta =0 \pmod{2}$,  is satisfied.  On the other hand,  due to (\ref{eq:property1/2}),  
 the choice $\al_{\rm R} =u$ is equivalent to $\al_{\rm R} =1-u$,  enabling also helices of reversed chirality (\ref{eq:SHSminus})
according to (\ref{eq:helixMinusCondition}), 
at anisotropies $(N+1)\eta = 1\pmod{2 }$, see bottom panel of Fig.~\ref{FigPurityB09}.

\section{Discussion}

We have presented analytic expression for a unique steady state of a boundary driven XYZ brickwork circuit, via a  spatially inhomogeneous Matrix Product Ansatz with formally infinite bond dimension.   
Our analysis reveals the existence of especially simple NESS types - pure factorized chiral qubit configurations (elliptic spin helices) and their descendants, helices with kinks.  We find conditions for the existence of helices both in open dissipative setup,  and for fully coherent periodic brickwork qubit system.
Helices are characterized by a periodic modulation of qubit polarization,  with wave length inversely proportional to  
the XYZ gate anisotropy $\eta$,  both for ``easy-plane-like" case (real $\eta$) and  for ``easy-axis-like" 
case (imaginary $\eta$).  Elliptic ``easy-plane"  helices reduce to conventional trigonometric ones 
with harmonic modulations of qubits' polarization in the XXZ limit.  Periodic ``easy-axis-like"  helices do not have 
trigonometric counterparts.  Our findings show that the bulk qubit dynamics in systems without $U(1)$ symmetry is amenable to analytic treatment. In addition it exhibits a rich family of robust, chiral, and easily creatable steady states -- elliptic helices which are potentially important for experimental applications.

\begin{acknowledgments}
X.Z. acknowledges financial support from the National
  Natural Science Foundation of China (No. 12575007).
 V.P.  acknowledges support by ERC Advanced grant
 No.~101096208 -- QUEST,  Research Program P1-0402 and Grant N1-0368 of Slovenian Research and Innovation Agency (ARIS) and by Deutsche Forschungsgemeinschaft through DFG project KL645/20-2. 
\end{acknowledgments}

\bibliography{bibfile}

\appendix
\begin{widetext}
\section{Theta functions}\label{App:theta}
We introduce the following Jacobi theta functions $\vartheta_{\al}(u,q)$ \cite{WatsonBook}
\begin{align}
\begin{aligned}
&\vartheta_{1}(u,q)=2\sum_{n=0}^\infty(-1)^n q^{(n+\frac12)^2}\sin[(2n+1)u],\\
&\vartheta_{2}(u,q)=2\sum_{n=0}^\infty q^{(n+\frac12)^2}\cos[(2n+1)u],\\
&\vartheta_{3}(u,q)=1+2\sum_{n=1}^\infty q^{n^2}\cos(2nu),\\
&\vartheta_{4}(u,q)=1+2\sum_{n=1}^\infty (-1)^nq^{n^2}\cos(2nu).
\end{aligned}
\end{align}
For convenience, we use the following shorthand notations $ \el{\al}(u),\,\bell{\al}(u),\,\alpha=1,2,3,4$
\begin{align}
&\el{\al}(u) \equiv  \vartheta_{\al} (\pi u,e^{2i\pi\tau}),\quad \bell{\al}(u) \equiv   \vartheta_{\al} (\pi u ,e^{i\pi\tau}),
\end{align}
where $\tau$ is a complex number with a positive imaginary part.

\section{Effective Krauss operators}
\label{app:KraussMatrices}

Here we show that  local boundary channels
\begin{align}
\begin{aligned}
\mathcal K_{\rm L}(\rho) &= \tr_1 U (\rho_{\rm L}\otimes \rho)U^\dagger,\\
\mathcal K_{\rm R}(\rho) &= \tr_2 U (\rho\otimes \rho_{\rm R})U^\dagger.  
\end{aligned}\label{app:reset}
\end{align}
with pure $\rho_{\rm L/R}= \ket{\psi_{\rm L/R}}  \bra{\psi_{\rm L/R}}$,  
 can be expressed  in terms of a pair of Krauss matrices $K^{[1]}_{\rm L/R},K^{[2]}_{\rm L/R}$
as 
$\mathcal K_{\rm L/R}(\rho) = \sum_{\mu=1}^2 K^{[\mu]\dagger}_{\rm L/R} \rho K^{[\mu]}_{\rm L/R}$.  
Note that we do not need an explicit form of the Krauss matrices $ K^{[\mu]}_{\rm L/R}$ for the proof of our MPA (\ref{eq:MPA}).
Here we give the derivation of the Krauss matrices   for completeness
and for possible use in numerical algorithms.   

Let us introduce a unitary matrix $V$ such that $\ket{\psi_{\rm L}} = V \binom {1}{0}$,  and 
define a trace-orthonormal basis of $2\times 2$ matrices $\vfi_j$ as
  \begin{align*}
&\vfi_0 =  V \frac{I+\si^z}{2} V^\dagger \equiv \ket{\psi_{\rm L}}  \bra{\psi_{\rm L}} = \rho_{\rm L},\\
&\vfi_1 =  V  \frac{I-\si^z}{2} V^\dagger,\\
&\vfi_2 =  V \si^+ V^\dagger,\\
&\vfi_3 =  V \si^{-} V^\dagger,
\end{align*}
satisfying 
\begin{align*}
&{\rm tr}(\vfi_j \vfi_k^\dagger) = \de_{j,k}.
\end{align*}
Using the basis, we can expand the XYZ gate $U$ from (\ref{eq:gateU}) as 
\begin{align*}
&U=\sum_{k=0}^3 \vfi_k \otimes u_k,\\
&U^\dagger=\sum_{k=0}^3 \vfi_k^\dagger \otimes u_k^\dagger,\\
&u_k= {\rm tr}_1((\vfi_k^\dagger \otimes I)U ).
\end{align*}
Now,  for arbitrary $2\times 2$ matrix $A$, we have 
\begin{align}
&\mathcal K_{\rm L}(A) = \tr_1 U (\rho_{\rm L}\otimes A)U^\dagger \no \\
&= \sum_{k,j=0}^3  \tr_1 (\vfi_k \otimes u_k) (\vfi_0\otimes A) (\vfi_j^\dagger \otimes u_j^\dagger) \no \\
&=  \sum_{k,j=0}^3 \tr(\vfi_k \vfi_0 \vfi_j^\dagger)  \, (u_k A u_j^\dagger)\no \\
&= \sum_{k,j=0}^3 C_{j,k}  \, (u_k A u_j^\dagger), \label{app:CjkSum}
\end{align}
where $C_{j,k} = \tr (\vfi_k \vfi_0 \vfi_j^\dagger) $. Only nonzero $C_{j,k}$ coefficients are 
$C_{0,0}=C_{3,3}=1$. Denoting $K^{[1]}_{\rm L} \equiv u_{0}$,  $K^{[2]}_{\rm L} \equiv u_{3}$, we finally obtain
\begin{align}
&\mathcal K_{\rm L}(A)  = \sum_{\mu=1}^2 K^{[\mu]}_{\rm L} A K^{[\mu]\dagger}_{\rm L},\\
&K^{[1]}_{\rm L}= \tr_1((\vfi_0^\dagger \otimes I)U ),\label{app:K1} \\
&K^{[2]}_{\rm L}= \tr_1((\vfi_3^\dagger \otimes I)U ).\label{app:K2} 
\end{align}
Taking $U$ in the form 
\begin{align}
&U(u,\eta)=\frac{1}{\sqrt{Y}}\left(
\begin{array}{cccc}
 r_1 & 0 & 0 & r_4 \\
 0 & r_3 & r_2 & 0 \\
 0 & r_2& r_3 & 0 \\
 r_4 & 0 & 0 & r_1 \\
\end{array}
\right),\\
&r_1= \frac{\el{4}(u)\el{1}(u+\eta)}{\el{4}(0)\el{1}(\eta)},\quad r_2=\frac{\el{1}(u) \el{4}(u+\eta)}{\el{4}(0)\el{1}(\eta)},\\
&r_3= \frac{\el{4}(u)\el{4}(u+\eta)}{\el{4}(0)\el{4}(\eta)},\quad r_4=\frac{\el{1}(u) \el{1}(u+\eta)}{\el{4}(0) \el{4}(\eta)},
\label{app:Ugate}\\
&Y=\frac{\bell{1}(\eta+u)\bell{1}(\eta-u)}{\bell{1}(\eta)\bell{1}(\eta)},\no
\end{align}
and parameterizing $\ket{\psi_{\rm L}}$ via a polar angle  $\th_{\rm L}$  and azimuthal angle $\vfi_{\rm L}$,  we have 
 \begin{align}
&\ket{\psi_{\rm L}} = \binom{e^{-i \vfi_{\rm L}/2} \cos \frac{\th_{\rm L}}{2}}{e^{i \vfi_{\rm L}/2} \sin \frac{\th_{\rm L}}{2}} =
V \binom{1}{0},\\
&V= \exp(-i \vfi_{\rm L} \tfrac{\si^z}{2})
\left(
\begin{array}{cc}
 \cos \frac{\th_{\rm L} }{2} & -\sin \frac{\th_{\rm L}}{2}\\[4pt]
 \sin \frac{\th_{\rm L} }{2} & \cos \frac{\th_{\rm L} }{2} \\
\end{array}
\right),
\end{align}
Calculating $K^{[1]}_{\rm L}$,  $K^{[2]}_{\rm L} $  from (\ref{app:K1}), (\ref{app:K2})  we obtain 
\begin{align}
K^{[1]}_{\rm L} &=\frac{1}{\sqrt{Y}}
\begin{pmatrix}
f_{\rm L}(r_1, r_3)& g_{\rm L}(r_2,r_4)  F_{\rm L}
\cr
g_{\rm L}(r_4,r_2)  F_{\rm L} &f_{\rm L}(r_3, r_1)
\end{pmatrix},\\
K^{[2]}_{\rm L} &=\frac{1}{\sqrt{Y}}
\begin{pmatrix}
(r_3-r_1)F_{\rm L}& \tilde{f}_{\rm L}( r_2,r_4)
\cr
\tilde{f}_{\rm L}(r_4,r_2) &  (r_1-r_3)F_{\rm L}
\end{pmatrix},\label{app:KraussL}
\end{align}
where 
\begin{align}
\begin{aligned}\label{app:defL} 
&F_{\rm L}= \frac{ \sin \th_{\rm L}}{2}, \\
&\tilde{f}_{\rm L}(x,y)= x  e^{-i \varphi_{\rm L}} \cos^2 \frac{\th_{ \rm L}}{2}  - y  e^{i \varphi_{\rm L}} \sin^2 \frac{\th_{\rm L}}{2},\\
&f_{\rm L}(x,y)= x  \cos^2 \frac{\th_{ \rm L}}{2}  +y  \sin^2 \frac{\th_{\rm L}}{2}, \\
&g_{\rm L}(x,y)=  e^{-i \varphi_{\rm L}} x+e^{i \varphi_{\rm L}}y.
\end{aligned}
\end{align}

Completely analogously we treat the right boundary.   We then obtain 
\begin{align}
&\mathcal K_{\rm R}(\rho)  = \tr_2 U (\rho\otimes \rho_{\rm R})U^\dagger= \sum_{\mu=1}^2 K^{[\mu]}_{\rm R} \rho K^{[\mu]\dagger}_{\rm R},
\end{align}
where $K^{[\mu]}_{\rm R}$ is given by exactly the same expressions as for 
$K^{[\mu]}_{\rm L}$,  with the substitutions of all
subscripts ${\rm L} \rightarrow {\rm R}$ in all the expressions in (\ref{app:KraussL}), (\ref{app:defL}).

\section{Proof of Eq. (\ref{eq:RLL}) } \label{App:A}

Let us introduce some useful identities \cite{Takhtajan:1979,Slavnov:2020}
\begin{align}
&\widetilde{U}\ket{\psi(x\mp u)}\otimes\ket{\psi(x\pm\eta)}=\frac{\bell{1}(u+\eta)}{\bell{1}(\eta)}\ket{\psi(x)}\otimes\ket{\psi(x\mp u\pm\eta)},
\label{App:A1}\\
&\widetilde{U}\ket{\psi(x\pm\eta)}\otimes\ket{\psi(y\mp u)}\no\\
&=\frac{\bell{2}(\frac{x+y}{2}\mp u)}{\bell{2}(\frac{x+y}{2})}\ket{\psi(x\pm u\pm\eta)}\otimes\ket{\psi(y)}+\frac{\bell{1}(u)\bell{2}(\frac{x+y}{2}\pm \eta)}{\bell{1}(\eta)\bell{2}(\frac{x+y}{2})}\ket{\psi(y\mp u\mp\eta)}\otimes\ket{\psi(x)},\label{App:A2}\\
&\bra{\xi(x\mp u)}\otimes\bra{\xi(x\pm\eta)}\widetilde{U}=\frac{\bell{1}(u+\eta)}{\bell{1}(\eta)}\bra{\psi(x)}\otimes\bra{\psi(x\mp u\pm\eta)},\label{App:A3}\\
&\bra{\xi(x\pm\eta)}\otimes\bra{\xi(y\mp u)}\widetilde{U}\no\\
&=\frac{\bell{2}(\frac{x+y}{2}\mp u)}{\bell{2}(\frac{x+y}{2})}\bra{\xi(x\pm u\pm\eta)}\otimes\bra{\xi(y)}+\frac{\bell{1}(u)\bell{2}(\frac{x+y}{2}\pm \eta)}{\bell{1}(\eta)\bell{2}(\frac{x+y}{2})}\bra{\xi(y\mp u\mp\eta)}\otimes\bra{\xi(x)}.\label{App:A4}
\end{align}

Due to band structure of the matrices (\ref{eq:Lax}), for  a proof  of Eq. (\ref{eq:RLL}) it is enough to check the following properties:
(below  we adopt simplified notations $L(m) \rightarrow L_m,  \, X(m)  \rightarrow X_m$ etc.) 
\begin{align}
\begin{aligned}\label{eq:RLL12}
&U(L_m \otimes L_{m+1}' )=  (L_m' \otimes L_{m+1}) U, \\
&U( L_m \otimes X_{m+1} + X_m' \otimes L_{m-1}') \\
&=( L_m'\otimes X_{m+1}' + X_m \otimes L_{m-1}) U,  \\
&U (X_m' \otimes X_{m-1}) =  (X_m \otimes X_{m-1}') U,
\end{aligned}
\end{align}
which hold for arbitrary $m$. And the same set for the conjugated operators in (\ref{eq:Lax}). 

Define the following operators
\begin{align}
\mc{L}_m&=\ket{\psi(m\eta+b)}\bra{\xi(-u-m\eta-b)},\quad \mc{L}'_m=\ket{\psi(u+m\eta+b)}\bra{\xi(-m\eta-b)}\\
\mc{X}_m&=\ket{\psi(m\eta-u+a)}\bra{\xi(-m\eta-a)},\quad 
\mc{X}'_m=\ket{\psi(m\eta+a)}\bra{\xi(u-m\eta-a)}.
\end{align}
By using Eqs. (\ref{App:A1})-(\ref{App:A4}) repeatedly, we can get the following identities 
\begin{align}
&\widetilde{U}(\mc{L}_m \otimes \mc{L}_{m+1}' )=  (\mc{L}_m' \otimes \mc{L}_{m+1})\widetilde{U},\\
&\widetilde{U}(\mc{X}_m' \otimes \mc{X}_{m-1}) =  (\mc{X}_m \otimes \mc{X}_{m-1}')\widetilde{U},\\
&\widetilde{U}\left( \bell{2}(\tfrac{a+b}{2}+m\eta-\eta)\mc{L}_m \otimes \mc{X}_{m+1} + \bell{2}(\tfrac{a+b}{2}+m\eta+\eta)\mc{X}_m' \otimes \mc{L}_{m-1}'\right)\no\\
&=\left( \bell{2}(\tfrac{a+b}{2}+m\eta-\eta)\mc{L}_m'\otimes \mc{X}_{m+1}' + \bell{2}(\tfrac{a+b}{2}+m\eta+\eta)\mc{X}_m \otimes \mc{L}_{m-1}\right)\widetilde{U}. 
\end{align}
We suppose
\begin{align}
L_m=\frac{\mc{L}_m}{\alpha_m},\quad L'_m=\frac{\mc{L}'_m}{\alpha'_m},\quad X_m=\frac{\mc{X}_m}{\beta_m},\quad X'_m=\frac{\mc{X}'_m}{\beta'_m}.
\end{align}
From Eq. \eqref{eq:RLL12}, the following constraints need to be satisfied 
\begin{align}
&\alpha_m\alpha'_{m+1}=\alpha'_m\alpha_{m+1},\qquad \beta_m\beta'_{m-1}=\beta'_m\beta_{m-1},\\
&\bell{2}(\tfrac{a+b}{2}+m\eta-\eta)\alpha_m\beta_{m+1}=\bell{2}(\tfrac{a+b}{2}+m\eta+\eta)\beta'_m\alpha'_{m-1}\no\\
&=\bell{2}(\tfrac{a+b}{2}+m\eta-\eta)\alpha'_m\beta'_{m+1}=\bell{2}(\tfrac{a+b}{2}+m\eta+\eta)\beta_m\alpha_{m-1}.
\end{align}
A simple solution is 
\begin{align}
&\alpha_m=\alpha'_m=\beta_m=\beta'_m=\bell{2}(\tfrac{a+b}{2}+m\eta).
\end{align}
Reinstalling the original notations 
($L_m \rightarrow L(m),  \, X_m  \rightarrow X(m)$ etc.)  and letting $b=a$,
we  obtain Eq. (\ref{eq:RLL}) in the main text.



\section{Resolving Eq. (\ref{eq:LBE}): Finding the Left Auxiliary Vector}\label{App:left}
We first introduce some identities regarding theta functions
\begin{align}
&\el{1}(u)\el{1}(v)+\el{4}(u)\el{4}(v)=\bell{4}\left(\frac{u+v}{2}\right)\bell{3}\left(\frac{u-v}{2}\right),\label{theta:eq:1}\\
&\bell{1}(u)\bell{4}(v)-\bell{1}(v)\bell{4}(u)=\el{2}\left(\frac{u+v}{2}\right)\el{1}\left(\frac{u-v}{2}\right),\label{theta:eq:3}\\
&\bell{1}(x) \bell{2}(y) \bell{3}(z)\bell{4}(w) -\bell{1}\left(\frac{x+y+z-w}{2}\right) \bell{2}\left(\frac{x+y-z+w}{2}\right) \bell{3}\left(\frac{x-y+z+w}{2}\right) \bell{4}\left(\frac{y+z+w-x}{2}\right)\no\\
&-\bell{1}\left(\frac{w+x-y-z}{2}\right) \bell{2}\left(\frac{w-x+y-z}{2}\right) \bell{3}\left(\frac{w-x-y+z}{2}\right) \bell{4}\left(\frac{w+x+y+z}{2}\right)=0.\label{theta:eq:2}
\end{align}
Another useful identities derived from Eqs. \eqref{App:A1} and \eqref{App:A2} are
\begin{align}
U\ket{\psi(\beta)}\otimes\ket{\psi(u+\beta+\eta)}&= \frac{\bell{1}(u+\eta)}{\bell{1}(\eta)}\ket{\psi(u+\beta)}\otimes\ket{\psi(\beta+\eta)},\label{U:psi:1} \\
U\ket{\psi(\be)}\otimes\ket{\psi(\be-\eta+u)}
&=\frac{1}{\kappa}\frac{\bell{2}(\be+u)}{\bell{2}(\be)}\ket{\psi(\be- u)}\otimes\ket{\psi(\be-\eta)}+\frac{1}{\kappa}\frac{\bell{1}(u)\bell{2}(\be- \eta)}{\bell{1}(\eta)\bell{2}(\be)}\ket{\psi(\be+ u)}\otimes\ket{\psi(\be+\eta)},\label{U:psi:2}\\
U\ket{\psi(\be)}\otimes\ket{\psi(\eta-u+\beta)}
&=\frac{1}{\kappa}\frac{\bell{2}(\beta- u)}{\bell{2}(\beta)}\ket{\psi(\beta+u)}\otimes\ket{\psi(\be+\eta)}+\frac{1}{\kappa}\frac{\bell{1}(u)\bell{2}(\beta+\eta)}{\bell{1}(\eta)\bell{2}(\beta)}\ket{\psi(\be-u)}\otimes\ket{\psi(\beta-\eta)}.\label{U:psi:3}
\end{align}

We postulate the solution of the Eq. (\ref{eq:LBE}) in the form 
\begin{equation}
\bra{\rm L} = \sum_{j,j'=0,1}l_{j,j'} \bra{j}_{\mathfrak{a}}\bra{j'}_{\mathfrak{b}}\,,\;\;
\label{eq:LeftAuxiliaryVec}
\end{equation}
for appropriately chosen parameter $a$ in all Eqs. (\ref{eq:LaxMatrices}).

First,  without losing generality, we postulate the left reset channel to target the qubit state 
\begin{align}
&\rho_{\rm L}=\rho_0=\frac{\ket{\psi(\be)} \bra{\psi(\be)}}{\braket{\psi(\be)}{\psi(\be)}},\quad 
\rho_{n}=\frac{\ket{\psi(\be+n\eta)} \bra{\psi(\be+n\eta)}}{\braket{\psi(\be+n\eta)}{\psi(\be+n\eta)}},\quad\rho'_{n}=\frac{\ket{\psi(u+\be+n\eta)} \bra{\psi(u+\be+n\eta)}}{\braket{\psi(u+\be+n\eta)}{\psi(u+\be+n\eta)}},
\end{align}
where $\be=\alpha_{\rm L}=w_0+iw_1$ is an arbitrary complex number.

Postulating the constants $a=\be$ in Eq.   (\ref{eq:LaxMatrices}), 
we write Eq. (\ref{eq:LBE}) in components
\begin{align}
&\sum_{j,j'=0,1} l_{j,j'}   (L^{+}_1)_{j,k}  (T^{+}_1)_{j',k'} = \sum_{j,j'=0,1}  l_{j,j'}   \mathcal{K}_{\rm L}[(L^{-}_1)_{j,k}  (T^{-}_1)_{j',k'}],\quad k,k'=0,1.
\label{eq(kkp)}
\end{align}

Before analyzing Eq.~\eqref{eq(kkp)}, we denote $L(m) \equiv L_m$, $L'(m) \equiv L'_m$, $X(m) \equiv X_m$, $X'(m) \equiv X'_m$, $L(m)^\dagger \equiv T_m$, $L'(m)^\dagger \equiv T'_m$, $X(m)^\dagger \equiv Y_m$, $X'(m)^\dagger \equiv Y'_m$, and their definitions are as follows
\begin{align}
&\begin{aligned}
&L_m=\frac{1}{C_m} \ket{\psi(m\eta+\be)}\bra{\xi(-u-m\eta-\be)}, \\
 &L'_m=\frac{1}{C_m} \ket{\psi(u+m\eta+\be)}\bra{\xi(-m\eta- \be)},\\
&X_m=\frac{1}{C_m} \ket{\psi(m\eta-u+\be)}\bra{\xi(-m\eta-\be)},\\
&X'_m=\frac{1}{C_m} \ket{\psi(m\eta+\be)}\bra{\xi(u-m\eta-\be)},
\end{aligned}\\
& C_m=\bell{2}(\beta+m\eta),
\end{align}
and 
\begin{align}
\begin{aligned}
&T_m=\frac{1}{C^*_m} \ket{\xi(-u-m\eta-\be)}\bra{\psi(m\eta+\be)},\\
&T'_m=\frac{1}{C^*_m} \ket{\xi(-m\eta- \be)}\bra{\psi(u+m\eta+\be)},\\
&Y_m=\frac{1}{C^*_m} \ket{\xi(-m\eta-\be)} \bra{\psi(m\eta-u+\be)},\\
&Y'_m=\frac{1}{C^*_m} \ket{\xi(u-m\eta-\be)} \bra{\psi(m\eta+\be)}.
\end{aligned}
\end{align}
\paragraph*{Expression of $l_{0,0}$}
Taking $(k,k')=(0,0)$ in Eq. \eqref{eq(kkp)} yields
\begin{align}
l_{0,0}L_1 T_1 &=l_{0,0}{\cal K}_{\rm L} [ L_1' T_1'].\label{eq(0,0)}
\end{align}
The definition of the $\mathcal{K}_{\rm L}$ is 
\begin{equation}
\mathcal K_{\rm L}(A) = \tr_1\left[U(\rho_{\rm L}\otimes A)U^\dagger\right].
\label{eq:resetL}
\end{equation}
By substituting $A=L'_1 T'_1$, we obtain 
\begin{align}
{\cal K}_{\rm L}[L'_1 T'_1] &= \frac{\bra{\xi(-\eta- \be)}\ket{\xi(-\eta- \be)}}{C_1C^*_1\braket{\psi(\be)}{\psi(\be)}} {\rm tr}_1\left(U \ket{\psi(\beta)}\otimes \psi(\beta+u+\eta)\bra{\psi(\beta)}\otimes \bra{\psi(\beta+u+\eta)} U^\dagger\right)\no\\
&\overset{\eqref{U:psi:1}}{=}\frac{\bra{\xi(-\eta- \be)}\ket{\xi(-\eta- \be)}}{C_1C^*_1\braket{\psi(\be)}{\psi(\be)}} \braket{\psi(\be+u)}{\psi(\be+u)}\ket{\psi(\eta+\beta)}\bra{\psi(\eta+\be)}.
\label{simpleKLaction}
\end{align}
In the both case (A) and case (B), we find
\begin{align}
\frac{\bra{\xi(-\eta- \be)}\ket{\xi(-\eta- \be)}\braket{\psi(\be+u)}{\psi(\be+u)}}{\bra{\xi(-u-\eta-\be)}\ket{\xi(-u-\eta-\be)}\braket{\psi(\be)}{\psi(\be)}}\overset{\eqref{theta:eq:1}}{=}1,
\end{align}
which directly leads to Eq. \eqref{eq(0,0)}. Since Eq. \eqref{eq(0,0)} holds for arbitrary $l_{0,0}$, we set $l_{0,0}=1$.


\paragraph*{Expression of $l_{0,1}$ and $l_{1,0}$}
For the  element $(k,k') = (0,1)$  of (\ref{eq(kkp)}) we get 
\begin{align}
&L_1 Y_1' + l_{0,1} L_1 T_{-1} = {\cal K}_{\rm L} [L_1'Y_1]+  l_{0,1} {\cal K}_{\rm L} [L'_1T_{-1}']. \label{eq(0,1)}
\end{align}
Using Eqs. \eqref{U:psi:1} - \eqref{U:psi:3} repeatedly,  we have
\begin{align}
\mathcal{K}_{\rm L}[L_1' Y_1]
&=\frac{\braket{\xi(-\eta-\be)}{\xi(-\eta-\be)}}{C_1C_{1}^* \braket{\psi(\be)}{\psi(\be)}}\frac{1}{\kappa}\frac{\bell{1}(u+\eta)}{\bell{1}(\eta)}\frac{1}{\kappa^*}\frac{\bell{2}(\beta^*- u^*)}{\bell{2}(\beta^*)}\braket{\psi(\be+u)}{\psi(\be+u)} \ket{\psi(\be+\eta)}\bra{\psi(\be+\eta)}\no\\
&\qquad+\frac{\braket{\xi(-\eta-\be)}{\xi(-\eta-\be)}}{C_1\braket{\psi(\be)}{\psi(\be)}}\frac{1}{\kappa}\frac{\bell{1}(u+\eta)}{\bell{1}(\eta)}\frac{1}{\kappa^*}\frac{\bell{1}(u^*)}{\bell{1}(\eta^*)\bell{2}(\beta^*)}\braket{\psi(\be-u)}{\psi(\be+u)}\ket{\psi(\be+\eta)}\bra{\psi(\be-\eta)},\\
\mathcal{K}_{\rm L}[L'_1T'_{-1}]&=\frac{\braket{\xi(-\eta-\be)}{\xi(\eta-\be)}}{C_1C_{-1}^*\braket{\psi(\be)}{\psi(\be)}} \frac{1}{\kappa}\frac{\bell{1}(u+\eta)}{\bell{1}(\eta)}\frac{1}{\kappa^*}\frac{\bell{2}(\be^*+u^*)}{\bell{2}(\be^*)}\braket{\psi(\be-u)}{\psi(\be+u)}\ket{\psi(\be+\eta)}\bra{\psi(\be-\eta)}\no\\
&\quad+\frac{\braket{\xi(-\eta-\be)}{\xi(\eta-\be)}}{C_1\braket{\psi(\be)}{\psi(\be)}} \frac{1}{\kappa}\frac{\bell{1}(u+\eta)}{\bell{1}(\eta)}\frac{1}{\kappa^*}\frac{\bell{1}(u^*)}{\bell{1}(\eta^*)\bell{2}(\be^*)}\braket{\psi(\be+u)}{\psi(\be+u)}\ket{\psi(\be+\eta)}\bra{\psi(\be+\eta)}.
\end{align}
Comparing the two sides of Eq.~\eqref{eq(0,1)} gives the following identities
\begin{align}
&\frac{\braket{\xi(-u-\eta-\be)}{\xi(u-\eta-\be)}}{C_1^*\braket{\psi(\be+u)}{\psi(\be+u)}}l_{0,0}-\frac{\braket{\xi(-\eta-\be)}{\xi(-\eta-\be)}}{C_{1}^* \braket{\psi(\be)}{\psi(\be)}}\frac{1}{\kappa}\frac{\bell{1}(u+\eta)}{\bell{1}(\eta)}\frac{1}{\kappa^*}\frac{\bell{2}(\beta^*- u^*)}{\bell{2}(\beta^*)}l_{0,0}\no\\
&-\frac{\braket{\xi(-\eta-\be)}{\xi(\eta-\be)}}{\braket{\psi(\be)}{\psi(\be)}} \frac{1}{\kappa}\frac{\bell{1}(u+\eta)}{\bell{1}(\eta)}\frac{1}{\kappa^*}\frac{\bell{1}(u^*)}{\bell{1}(\eta^*)\bell{2}(\be^*)}l_{0,1}=0,\label{eq:l01:1}\\
&\frac{\braket{\xi(-u-\eta-\be)}{\xi(-u+\eta-\be)}}{C_{-1}^*\braket{\psi(\be-u)}{\psi(\be+u)}}l_{0,1}-\frac{\braket{\xi(-\eta-\be)}{\xi(-\eta-\be)}}{\braket{\psi(\be)}{\psi(\be)}}\frac{1}{\kappa}\frac{\bell{1}(u+\eta)}{\bell{1}(\eta)}\frac{1}{\kappa^*}\frac{\bell{1}(u^*)}{\bell{1}(\eta^*)\bell{2}(\beta^*)}l_{0,0}\no\\
&-\frac{\braket{\xi(-\eta-\be)}{\xi(\eta-\be)}}{C_{-1}^*\braket{\psi(\be)}{\psi(\be)}} \frac{1}{\kappa}\frac{\bell{1}(u+\eta)}{\bell{1}(\eta)}\frac{1}{\kappa^*}\frac{\bell{2}(\be^*+u^*)}{\bell{2}(\be^*)}l_{0,1}=0.\label{eq:l01:2}
\end{align}
From Eq. \eqref{eq:l01:1}, we can derive the expression of $l_{0,1}$, specially as follows 
\begin{align}
\mbox{case (A)}:\quad l_{0,1}&
\overset{(\ref{theta:eq:1})}{=}\frac{ \bell{1}(\eta ) \bell{2}(\beta^*-u) \bell{3}(iw_1+\eta ) \bell{4}(w_0)}{\bell{1}(u)\bell{2}(\beta^*-\eta) \bell{3}(iw_1) \bell{4}(w_0+\eta )}-\frac{\bell{1}(\eta -u)\bell{2}(\beta^*) \bell{3}(iw_1+\eta +u)\bell{4}(w_0)\bell{4}(w_0) }{\bell{1}(u)\bell{2}(\beta^*-\eta )\bell{3}(iw_1) \bell{4}(w_0+\eta ) \bell{4}(w_0+u)}\no\\
&\overset{(\ref{theta:eq:2})}{=}\frac{\bell{4}(w_0)\bell{3}(u+iw_1)}{\bell{4}(u+w_0)\bell{3}(iw_1)},\label{exp:l01:A}\\
\mbox{case (B)}:\quad l_{0,1}
&\overset{(\ref{theta:eq:1})}{=}\frac{\bell{1}(\eta)\bell{2}(u+\be^*)\bell{3}(iw_1)\bell{4}(w_0+\eta)}{\bell{1}(u)\bell{2}(\be^*+\eta)\bell{3}(iw_1+\eta)\bell{4}(w_0)}-\frac{\bell{1}(\eta-u)\bell{2}(\be^*)\bell{3}(iw_1)\bell{3}(iw_1)\bell{4}(u+\eta+w_0)}{\bell{1}(u)\bell{2}(\be^*+\eta)\bell{3}(iw_1+\eta)\bell{3}(u+iw_1)\bell{4}(w_0)}\no\\
&\overset{(\ref{theta:eq:2})}{=}\frac{\bell{4}(u+w_0)\bell{3}(iw_1)}{\bell{4}(w_0)\bell{3}(u+iw_1)}.\label{exp:l01:B}
\end{align}
One can verify that $l_{0,1}$ in \eqref{exp:l01:A} or \eqref{exp:l01:B} also satisfies Eq.~\eqref{eq:l01:2}.

Analogously, taking $(k,k') = (1,0)$ in (\ref{eq(kkp)}) can give $l_{1,0}$. After some tedious calculations, we finally obtain 
\begin{align}
&l_{1,0} =l_{0,1}^*.
\end{align}

\paragraph*{Expression of $l_{1,1}$} Setting $(k,k') = (1,0)$ in \eqref{eq(kkp)} yields the following equation
\begin{align}
&l_{0,0} X_1'Y_1'+l_{1,0}L_{-1}Y_1'+l_{0,1}X'_1T_{-1}+l_{1,1}L_{-1}T_{-1}\no\\
=&\,l_{0,0}\mathcal{K}_{\rm L}[X_1Y_1]+l_{1,0}\mathcal{K}_{\rm L}[L'_{-1}Y_1]+l_{0,1}\mathcal{K}_{\rm L}[X_1T'_{-1}]+l_{1,1}\mathcal{K}_{\rm L}[L'_{-1}T'_{-1}].\label{eq(1,1)}
\end{align}
With the help of Eqs. \eqref{U:psi:1} - \eqref{U:psi:3}, we arrive at the following equations
\begin{align}
\mathcal{K}_{\rm L}[X_1Y_1]
&=\frac{\braket{\xi(-\eta-\beta)}{\xi(-\eta-\beta)}}{C_1C_1^*\braket{\psi(\be)}{\psi(\be)}}\frac{1}{\kappa}\frac{\bell{2}(\beta-u)}{\bell{2}(\beta)}\frac{1}{\kappa^*}\frac{\bell{2}(\beta^*-u^*)}{\bell{2}(\beta^*)}\braket{\psi(\be+u)}{\psi(\be+u)}\ket{\psi(\be+\eta)} \bra{\psi(\be+\eta)}\no\\
&\quad+\frac{\braket{\xi(-\eta-\beta)}{\xi(-\eta-\beta)}}{C_1\braket{\psi(\be)}{\psi(\be)}}\frac{1}{\kappa}\frac{\bell{2}(\beta-u)}{\bell{2}(\beta)}\frac{1}{\kappa^*}\frac{\bell{1}(u^*)}{\bell{1}(\eta^*)\bell{2}(\be^*)}\braket{\psi(\be-u)}{\psi(\be+u)}\ket{\psi(\be+\eta)}\bra{\psi(\be-\eta)}\no\\
&\quad +\frac{\braket{\xi(-\eta-\beta)}{\xi(-\eta-\beta)}}{C_1^*\braket{\psi(\be)}{\psi(\be)}}\frac{1}{\kappa}\frac{\bell{1}(u)}{\bell{1}(\eta)\bell{2}(\beta)}\frac{1}{\kappa^*}\frac{\bell{2}(\be^*-u^*)}{\bell{2}(\be^*)}\braket{\psi(\be+u)}{\psi(\be-u)}\ket{\psi(\be-\eta)}\bra{\psi(\be+\eta)}\no\\
&\quad+\frac{\braket{\xi(-\eta-\beta)}{\xi(-\eta-\beta)}}{\braket{\psi(\be)}{\psi(\be)}}\frac{1}{\kappa}\frac{\bell{1}(u)}{\bell{1}(\eta)\bell{2}(\beta)}\frac{1}{\kappa^*}\frac{\bell{1}(u^*)}{\bell{1}(\eta^*)\bell{2}(\beta^*)}\braket{\psi(\be-u)}{\psi(\be-u)}\ket{\psi(\be-\eta)}\bra{\psi(\be-\eta)},
\end{align}
\begin{align}
\mathcal{K}_{\rm L}[L'_{-1}Y_1]
&=\frac{\bra{\xi(\eta- \be)}\ket{\xi(-\eta-\be)}}{C_{-1}C_1^*\braket{\psi(\be)}{\psi(\be)}}\frac{1}{\kappa}\frac{\bell{2}(\be+u)}{\bell{2}(\be)}\frac{1}{\kappa^*}\frac{\bell{2}(\be^*-u^*)}{\bell{2}(\be^*)}\braket{\psi(\be+u)}{\psi(\be-u)}\ket{\psi(\be-\eta)}\bra{\psi(\be+\eta)}\no\\
&\quad+\frac{\bra{\xi(\eta- \be)}\ket{\xi(-\eta-\be)}}{C_{-1}\braket{\psi(\be)}{\psi(\be)}}\frac{1}{\kappa}\frac{\bell{2}(\be+u)}{\bell{2}(\be)}\frac{1}{\kappa^*}\frac{\bell{1}(u^*)}{\bell{1}(\eta^*)\bell{2}(\be^*)}\braket{\psi(\be-u)}{\psi(\be-u)}\ket{\psi(\be-\eta)}\bra{\psi(\be-\eta)}\no\\
&\quad +\frac{\bra{\xi(\eta- \be)}\ket{\xi(-\eta-\be)}}{C_1^*\braket{\psi(\be)}{\psi(\be)}}\frac{1}{\kappa}\frac{\bell{1}(u)}{\bell{1}(\eta)\bell{2}(\be)}\frac{1}{\kappa^*}\frac{\bell{2}(\be^*-u^*)}{\bell{2}(\be^*)}\braket{\psi(\be+u)}{\psi(\be+u)}\ket{\psi(\be+\eta)}\bra{\psi(\be+\eta)}\no\\
&\quad +\frac{\bra{\xi(\eta- \be)}\ket{\xi(-\eta-\be)}}{\braket{\psi(\be)}{\psi(\be)}}\frac{1}{\kappa}\frac{\bell{1}(u)}{\bell{1}(\eta)\bell{2}(\be)}\frac{1}{\kappa^*}\frac{\bell{1}(u^*)}{\bell{1}(\eta^*)\bell{2}(\be^*)}\braket{\psi(\be-u)}{\psi(\be+u)}\ket{\psi(\be+\eta)}\bra{\psi(\be-\eta)},
\end{align}
\begin{align}
\mathcal{K}_{\rm L}[X_1T'_{-1}]&=\frac{\bra{\xi(-\eta- \be)}\ket{\xi(\eta-\be)}}{C_{1}C_{-1}^*\braket{\psi(\be)}{\psi(\be)}}\frac{1}{\kappa}\frac{\bell{2}(\be-u)}{\bell{2}(\be)}\frac{1}{\kappa^*}\frac{\bell{2}(\be^*+u^*)}{\bell{2}(\be^*)}\braket{\psi(\be-u)}{\psi(\be+u)}\ket{\psi(\be+\eta)}\bra{\psi(\be-\eta)}\no\\
&\quad+\frac{\bra{\xi(-\eta- \be)}\ket{\xi(\eta-\be)}}{C_1\braket{\psi(\be)}{\psi(\be)}}\frac{1}{\kappa}\frac{\bell{2}(\be-u)}{\bell{2}(\be)}\frac{1}{\kappa^*}\frac{\bell{1}(u^*)}{\bell{1}(\eta^*)\bell{2}(\be^*)}\braket{\psi(\be+u)}{\psi(\be+u)}\ket{\psi(\be+\eta)}\bra{\psi(\be+\eta)}\no\\
&\quad +\frac{\bra{\xi(-\eta-\be)}\ket{\xi(\eta-\be)}}{C_{-1}^*\braket{\psi(\be)}{\psi(\be)}}\frac{1}{\kappa}\frac{\bell{1}(u)}{\bell{1}(\eta)\bell{2}(\be)}\frac{1}{\kappa^*}\frac{\bell{2}(\be^*+u^*)}{\bell{2}(\be^*)}\braket{\psi(\be-u)}{\psi(\be-u)}\ket{\psi(\be-\eta)}\bra{\psi(\be-\eta)}\no\\
&\quad +\frac{\bra{\xi(-\eta-\be)}\ket{\xi(\eta-\be)}}{\braket{\psi(\be)}{\psi(\be)}}\frac{1}{\kappa}\frac{\bell{1}(u)}{\bell{1}(\eta)\bell{2}(\be)}\frac{1}{\kappa^*}\frac{\bell{1}(u^*)}{\bell{1}(\eta^*)\bell{2}(\be^*)}\braket{\psi(\be+u)}{\psi(\be-u)}\ket{\psi(\be-\eta)}\bra{\psi(\be+\eta)},
\end{align}
\begin{align}
\mathcal{K}_{\rm L}[L'_{-1}T'_{-1}]&=\frac{\braket{\xi(\eta-\beta)}{\xi(\eta-\beta)}}{C_{-1}C_{-1}^*\braket{\psi(\be)}{\psi(\be)}}\frac{1}{\kappa}\frac{\bell{2}(\beta+u)}{\bell{2}(\beta)}\frac{1}{\kappa^*}\frac{\bell{2}(\beta^*+u^*)}{\bell{2}(\beta^*)}\braket{\psi(\be-u)}{\psi(\be-u)}\ket{\psi(\be-\eta)} \bra{\psi(\be-\eta)}\no\\
&\quad+\frac{\braket{\xi(\eta-\beta)}{\xi(\eta-\beta)}}{C_{-1}\braket{\psi(\be)}{\psi(\be)}}\frac{1}{\kappa}\frac{\bell{2}(\beta+u)}{\bell{2}(\beta)}\frac{1}{\kappa^*}\frac{\bell{1}(u^*)}{\bell{1}(\eta^*)\bell{2}(\be^*)}\braket{\psi(\be+u)}{\psi(\be-u)}\ket{\psi(\be-\eta)}\bra{\psi(\be+\eta)}\no\\
&\quad +\frac{\braket{\xi(\eta-\beta)}{\xi(\eta-\beta)}}{C_{-1}^*\braket{\psi(\be)}{\psi(\be)}}\frac{1}{\kappa}\frac{\bell{1}(u)}{\bell{1}(\eta)\bell{2}(\beta)}\frac{1}{\kappa^*}\frac{\bell{2}(\be^*+u^*)}{\bell{2}(\be^*)}\braket{\psi(\be-u)}{\psi(\be+u)}\ket{\psi(\be+\eta)}\bra{\psi(\be-\eta)}\no\\
&\quad+\frac{\braket{\xi(\eta-\beta)}{\xi(\eta-\beta)}}{\braket{\psi(\be)}{\psi(\be)}}\frac{1}{\kappa}\frac{\bell{1}(u)}{\bell{1}(\eta)\bell{2}(\beta)}\frac{1}{\kappa^*}\frac{\bell{1}(u^*)}{\bell{1}(\eta^*)\bell{2}(\beta^*)}\braket{\psi(\be+u)}{\psi(\be+u)}\ket{\psi(\be+\eta)}\bra{\psi(\be+\eta)}.
\end{align}
We see that both sides of Eq. \eqref{eq(1,1)} are linear combinations of $\ket{\psi(\beta+m_1\eta)}\bra{\psi(\beta+m_2\eta)}$, $m_1,m_2 = \pm 1$. Comparing the coefficients of $\ket{\psi(\beta+\eta)}\bra{\psi(\beta+\eta)}$ gives
\begin{align}
&\frac{\braket{\xi(u-\eta-\be)}{\xi(u-\eta-\be)}}{C_1C_1^*\braket{\psi(\beta+u)}{\psi(\be+u)}}l_{0,0}\no\\
&=l_{0,0}\frac{\braket{\xi(-\eta-\beta)}{\xi(-\eta-\beta)}}{C_1C_1^*\braket{\psi(\be)}{\psi(\be)}}\frac{1}{\kappa}\frac{\bell{2}(\beta-u)}{\bell{2}(\beta)}\frac{1}{\kappa^*}\frac{\bell{2}(\beta^*-u^*)}{\bell{2}(\beta^*)}\no\\
&\quad +l_{1,0}\frac{\bra{\xi(\eta- \be)}\ket{\xi(-\eta-\be)}}{C_1^*\braket{\psi(\be)}{\psi(\be)}}\frac{1}{\kappa}\frac{\bell{1}(u)}{\bell{1}(\eta)\bell{2}(\be)}\frac{1}{\kappa^*}\frac{\bell{2}(\be^*-u^*)}{\bell{2}(\be^*)}\no\\
&\quad+l_{0,1}\frac{\bra{\xi(-\eta- \be)}\ket{\xi(\eta-\be)}}{C_1\braket{\psi(\be)}{\psi(\be)}}\frac{1}{\kappa}\frac{\bell{2}(\be-u)}{\bell{2}(\be)}\frac{1}{\kappa^*}\frac{\bell{1}(u^*)}{\bell{1}(\eta^*)\bell{2}(\be^*)}\no\\
&\quad +l_{1,1}\frac{\braket{\xi(\eta-\beta)}{\xi(\eta-\beta)}}{\braket{\psi(\be)}{\psi(\be)}}\frac{1}{\kappa}\frac{\bell{1}(u)}{\bell{1}(\eta)\bell{2}(\beta)}\frac{1}{\kappa^*}\frac{\bell{1}(u^*)}{\bell{1}(\eta^*)\bell{2}(\beta^*)}.\label{eq:l11:1}
\end{align}
By substituting the expression of $l_{0,0}$ $l_{0,1}$ and $l_{1,0}$ into Eq. \eqref{eq:l11:1}, we can get 
\begin{align}
\mbox{case (A)}:\quad l_{1,1}&\overset{\eqref{theta:eq:1}}{=}-\frac{\bell{1}(\eta ) \bell{2}(\beta -u) \bell{3}\left(iw_1+u\right)\bell{4}\left(w_0+\eta \right) }{\bell{1}(u) \bell{2}(\beta +\eta ) \bell{3}\left(iw_1-\eta \right) \bell{4}\left(w_0+u\right)}+\frac{ \bell{1}(\eta )\bell{2}(\be^*-u)  \bell{3}\left(u-iw_1\right)\bell{4}\left(\eta -w_0\right)}{\bell{1}(u) \bell{2}(\be^*-\eta ) \bell{3}\left(iw_1-\eta \right) \bell{4}\left(w_0+u\right)}\no\\
&\qquad-\frac{\bell{1}(\eta -u) \bell{1}(\eta +u)\bell{2}(\beta ) \bell{2}(\be^*) \bell{3}\left(iw_1+\eta \right) \bell{4}\left(u-w_0\right)}{\bell{1}(u) \bell{1}(u) \bell{2}(\beta +\eta ) \bell{2}(\be^*-\eta )\bell{3}\left(iw_1-\eta \right) \bell{4}\left(w_0+u\right)}\no\\
&\qquad +\frac{\bell{1}(\eta ) \bell{1}(\eta ) \bell{2}(\beta -u) \bell{2}(\be^*-u)\bell{3}\left(iw_1+\eta \right)}{\bell{1}(u) \bell{1}(u)\bell{2}(\beta +\eta ) \bell{2}(\be^*-\eta )\bell{3}\left(iw_1-\eta \right)}\no\\
&\overset{\eqref{theta:eq:2}}{=}\frac{\bell{4}\left(u-w_0\right)}{\bell{4}\left(u+w_0\right)},\label{exp:l11:A}\\
\mbox{case (B)}:\quad l_{1,1}&\overset{\eqref{theta:eq:1}}{=}\frac{\bell{1}(\eta ) \bell{2}(\beta^*+u) \bell{3}\left(\eta -i w_1\right) \bell{4}\left(u-w_0\right)}{\bell{1}(u) \bell{2}(\beta^*+\eta ) \bell{3}\left(u+i w_1\right)\bell{4}\left(w_0-\eta \right) }+\frac{\bell{1}(\eta) \bell{1}(\eta) \bell{2}(\beta -u) \bell{2}(\beta^*+u)\bell{4}\left(w_0+\eta \right) }{\bell{1}(u) \bell{1}(u) \bell{2}(\beta +\eta ) \bell{2}(\beta^*+\eta ) \bell{4}\left(w_0-\eta \right)}\no\\
&\qquad-\frac{\bell{1}(\eta -u) \bell{1}(\eta +u)\bell{2}(\beta ) \bell{2}(\beta^*) \bell{3}\left(u-i w_1\right) \bell{4}\left(w_0+\eta \right) }{\bell{1}(u) \bell{1}(u) \bell{2}(\beta +\eta ) \bell{2}(\beta^*+\eta) \bell{3}\left(u+i w_1\right)\bell{4}\left(w_0-\eta \right) } \no\\
&\qquad -\frac{\bell{1}(\eta)\bell{2}(\beta -u) \bell{3}\left(\eta +i w_1\right) \bell{4}\left(w_0+u\right)}{\bell{1}(u) \bell{2}(\beta +\eta ) \bell{3}\left(u+i w_1\right)\bell{4}\left(w_0-\eta \right) }\no\\
&\overset{\eqref{theta:eq:2}}{=}\frac{\bell{3}(u-iw_1)}{\bell{3}(u+iw_1)}.\label{exp:l11:B}
\end{align}
Although Eq. \eqref{eq(1,1)} is overdetermined, comparing the coefficients of $\ket{\psi(\beta-\eta)}\bra{\psi(\beta+\eta)}$, $\ket{\psi(\beta+\eta)}\bra{\psi(\beta-\eta)}$, and $\ket{\psi(\beta-\eta)}\bra{\psi(\beta-\eta)}$ in Eq. \eqref{eq(1,1)} yields the same expression for $l_{1,1}$, namely Eqs. \eqref{exp:l11:A} or \eqref{exp:l11:B}.

\section{Resolving Eq. (\ref{eq:RBE}): Finding the right auxiliary vector  $\ket{\rm R}$. } \label{App:right}
Let us introduce the following identities derived from Eq. \eqref{App:A2}
\begin{align}
U\ket{\psi(k\eta+\beta)}\otimes\ket{\psi(\delta)}
&=\frac{\bell{2}(\frac{\delta+\eta+k\eta+\beta+u}{2})}{\kappa\,\bell{2}(\frac{\delta+\eta+k\eta+\beta-u}{2})}\ket{\psi(\beta+k\eta-u)}\otimes\ket{\psi(\delta-u)}\no\\
&\quad +\frac{\bell{1}(u)\bell{2}(\frac{\delta-\eta+k\eta+\beta-u}{2})}{\kappa\,\bell{1}(\eta)\bell{2}(\frac{\delta+\eta+k\eta+\beta-u}{2})}\ket{\psi(\delta+\eta)}\otimes\ket{\psi(k\eta+\beta+\eta)},\label{U:psi:4}\\
U\ket{\psi(k\eta+\beta)}\otimes\ket{\psi(\delta)}
&=\frac{\bell{2}(\frac{\delta-\eta+k\eta+\beta-u}{2})}{\kappa\,\bell{2}(\frac{\delta-\eta+k\eta+\beta+u}{2})}\ket{\psi(\beta+k\eta+u)}\otimes\ket{\psi(\delta+u)}\no\\
&\quad +\frac{\bell{1}(u)\bell{2}(\frac{\delta+\eta+k\eta+\beta+u}{2})}{\kappa\,\bell{1}(\eta)\bell{2}(\frac{\delta-\eta+k\eta+\beta+u}{2})}\ket{\psi(\delta-\eta)}\otimes\ket{\psi(k\eta+\beta-\eta)}.\label{U:psi:5}
\end{align}
One can write Eq. (\ref{eq:RBE}) in components.  
The element $(k,k')$ of the  Eq. (\ref{eq:RBE})  is
\begin{align}
& \sum_{j,j'=0}^{\infty} r_{j,j'}   (L^{-}_N)_{k,j}  (T^{-}_N)_{k',j'} = \sum_{j,j'=0}^{\infty}  r_{j,j'}   \mathcal K_{\rm R}[(L^{+}_N)_{k,j}  (T^{+}_N)_{k',j'}].
\label{eq(kkpR)}
\end{align}
The left-hand side of \eqref{eq(kkpR)} can be rewritten as
\begin{align}
\mbox{LHS}&=r_{k,k'}(L^{-}_N)_{k,k}  (T^{-}_N)_{k',k'}+ r_{k,k'+1}   (L^{-}_N)_{k,k}  (T^{-}_N)_{k',k'+1}+ r_{k+1,k'}   (L^{-}_N)_{k,k+1}  (T^{-}_N)_{k',k'}\no\\
&\quad + r_{k+1,k'+1}   (L^{-}_N)_{k,k+1}  (T^{-}_N)_{k',k'+1}\no\\
&=r_{k,k'}L'_{N-2k}T'_{N-2k'}+r_{k,k'+1}L'_{N-2k}Y_{N-2k'}+r_{k+1,k'}X_{N-2k}T'_{N-2k'}+r_{k+1,k'+1}X_{N-2k}Y_{N-2k'}\no\\
&=\frac{1}{C_{\bar k}C^*_{\bar k'}}\braket{\xi(-\bar k\eta-\be)}{\xi(-\bar k'\eta-\be)}\left[r_{k,k'}\ket{\psi(u+\bar k\eta+\be)}\bra{\psi(u+\bar k'\eta+\be)}\right.\no\\
&\quad+r_{k,k'+1}\ket{\psi(u+\bar k\eta+\be)}\bra{\psi(\bar k'\eta-u+\be)}+r_{k+1,k'}\ket{\psi(\bar k\eta-u+\be)}\bra{\psi(u+\bar k'\eta+\be)}\no\\
&\quad\left. +r_{k+1,k'+1}\ket{\psi(\bar k\eta-u+\be)}\bra{\psi(\bar k'\eta-u+\be)}\right],\label{R:LHS}
\end{align}
where $\bar k=N-2k$, $\bar k'=N-2k'$ and $\beta=\alpha_{\rm L}$.
The right-hand side of \eqref{eq(kkpR)} reads 
\begin{align}
\mbox{RHS}&=r_{k,k'}   \mathcal K_{\rm R}[(L^{+}_N)_{k,k}  (T^{+}_N)_{k',k'}]+r_{k,k'+1}   \mathcal K_{\rm R}[(L^{+}_N)_{k,k}  (T^{+}_N)_{k',k'+1}]\no\\
&\quad +r_{k+1,k'}   \mathcal K_{\rm R}[(L^{+}_N)_{k,k+1}  (T^{+}_N)_{k',k'}]+r_{k+1,k'+1}   \mathcal K_{\rm R}[(L^{+}_N)_{k,k+1}  (T^{+}_N)_{k',k'+1}]\no\\
&=r_{k,k'}   \mathcal K_{\rm R}[L_{N-2k}T_{N-2k'}]+r_{k,k'+1}   \mathcal K_{\rm R}[L_{N-2k}Y'_{N-2k'}]\no\\
&\quad +r_{k+1,k'}   \mathcal K_{\rm R}[X'_{N-2k} T_{N-2k'}]+r_{k+1,k'+1}   \mathcal K_{\rm R}[X'_{N-2k}Y'_{N-2k'}]\no\\
&=\frac{1}{C_{\bar k}C^*_{\bar k'}}\left[r_{k,k'}\braket{\xi(-u-\bar k\eta-\beta)}{\xi(-u-\bar k'\eta-\beta)}+r_{k,k'+1}\braket{\xi(-u-\bar k\eta-\beta)}{\xi(u-\bar k'\eta-\beta)}\right.\no\\
&\quad+\left.r_{k+1,k'}\braket{\xi(u-\bar k\eta-\beta)}{\xi(-u-\bar k'\eta-\beta)}+r_{k+1,k'+1}\braket{\xi(u-\bar k\eta-\beta)}{\xi(u-\bar k'\eta-\beta)}\right]\no\\
&\quad\times \mathcal K_{\rm R}\left[\ket{\psi(\bar k\eta+\beta)}\bra{\psi(\bar k'\eta+\beta)}\right].\label{R:RHS}
\end{align}
In the following, we will demonstrate that our ansatz in the main text (Eqs. \eqref{eq:AuxiliaryVecR}-\eqref{eq:nu} and \eqref{eq:CnmCaseA}) satisfies \eqref{eq(kkpR)} for arbitrary $k$ and $k'$.
For convenience, we will omit the overall factor $\frac{r_{k,k'}}{C_{\bar k}C^*_{\bar k'}}$ in the proof.

\paragraph*{Case (B)} To be consistent with the notation in the main text, we first rewrote $\alpha_{\rm L,R}$ as
\begin{align}
\alpha_{\rm L}=\beta=\Ga-N\eta-\eta,\quad \alpha_{\rm R}=\delta=u+\Ga+2\nu.
\end{align}
Multiplying Eq. \eqref{R:LHS} from the left by $\bra{\xi(\bar k\eta+\beta+u)}$ and from the right by $\ket{\xi(\bar k'\eta+\beta+u)}$ and using Eqs. \eqref{theta:eq:1}, \eqref{theta:eq:3}, the LHS becomes
\begin{align}
&\bra{\xi(\bar k\eta+\beta+u)}{{\rm LHS}}\ket{\xi(\bar k'\eta+\beta+u)}\no\\
&=\frac{r_{k+1,k'+1}}{r_{k,k'}}\braket{\xi(-\bar k\eta-\be)}{\xi(-\bar k'\eta-\be)}\bra{\xi(\bar k\eta+\beta+u)}\ket{\psi(\bar k\eta-u+\be)}\bra{\psi(\bar k'\eta-u+\be)}\ket{\xi(\bar k'\eta+\beta+u)}\no\\
&=-\frac{\bell{1}^2(u) \bell{1}(k\eta +\nu) \bell{1}(k'\eta +\nu^*)\bell{2}(\Ga-(2k+1)\eta  )\bell{2}(\Ga^*-(2k'+1)\eta)\bell{2}(\Ga- k\eta+\nu+u) \bell{2}(\Ga^*- k'\eta+\nu^*-u) }
{\bell{1}((k+1)\eta+\nu+u)\bell{1}((k'+1)\eta+\nu^*-u)\bell{2}(\Ga-(k+1)\eta+\nu) \bell{2}(\Ga^*-(k'+1)\eta+\nu^*)}\no\\
&\quad\times \frac{\bell{3}(i \Ga_1-(k-k')\eta ) \bell{4}\left(\Ga_0-(k+k'+1)\eta\right)\bell{4}((k+k'+2)\eta-\Ga_0) }{\bell{4}((k+k')\eta-\Ga_0)}.\label{R:LHS:exp}
\end{align}
The c-term in Eq. \eqref{R:RHS}, appearing in the first and second lines of the last step, can be simplified as
\begin{align}
&\quad \braket{\xi(-u-\bar k\eta-\beta)}{\xi(-u-\bar k'\eta-\beta)}+\frac{r_{k,k'+1}}{r_{k,k'}}\braket{\xi(-u-\bar k\eta-\beta)}{\xi(u-\bar k'\eta-\beta)}\no\\
&\quad+\frac{r_{k+1,k'}}{r_{k,k'}}\braket{\xi(u-\bar k\eta-\beta)}{\xi(-u-\bar k'\eta-\beta)}+\frac{r_{k+1,k'+1}}{r_{k,k'}}\braket{\xi(u-\bar k\eta-\beta)}{\xi(u-\bar k'\eta-\beta)}\no\\
&\overset{\eqref{theta:eq:1}}{=}\bell{4}\left(\Ga_0-(k+k'+1)\eta\right) \bell{3}\left(u+i \Ga_1- (k-k')\eta\right)+\frac{r_{k,k'+1}}{r_{k,k'}}\bell{4}\left(u+\Ga_0-(k+k'+1)\eta\right) \bell{3}\left(i \Ga_1- (k-k')\eta\right)\no\\
&\quad+\frac{r_{k+1,k'}}{r_{k,k'}}\bell{4}\left(u-\Ga_0+(k+k'+1)\eta\right) \bell{3}\left(i \Ga_1- (k-k')\eta\right)\no\\
&\quad +\frac{r_{k+1,k'+1}}{r_{k,k'}}\bell{4}\left(\Ga_0-(k+k'+1)\eta\right) \bell{3}\left(u-i \Ga_1+ (k-k')\eta\right)\no\\
&\overset{\eqref{theta:eq:2}}{=}\frac{\bell{1}(\eta +u) \bell{1}(\eta -u)  \bell{2}(\Gamma -(2 k+1)\eta) \bell{2}(\Ga^*-(2 k' +1)\eta) }{\bell{1}((k+1)\eta +\nu +u) \bell{1}((k' +1)\eta  +\nu^*-u)\bell{2}(\Gamma -(k+1)\eta  +\nu ) \bell{2}(\Ga^*-(k' +1)\eta  +\nu^*) }\no\\
&\quad \times\frac{ \bell{3}(i\Ga_1+\nu -\nu^*+u)\bell{4}(\Ga_0-(k+k'+1)\eta )\bell{4}(\Ga_0+\nu +\nu^*)}{\bell{4}( (k+k' )\eta -\Ga_0)}.\label{R:RHS:exp:1}
\end{align}
Since $\braket{\xi(x)}{\psi(x)}=0$, multiplying the third line of the last step in Eq. \eqref{R:RHS} on the left by $\bra{\xi(\bar k\eta+\beta+u)}$ and on the right by $\ket{\xi(\bar k'\eta+\beta+u)}$, and applying Eqs. \eqref{theta:eq:1}, \eqref{theta:eq:3} and \eqref{U:psi:5}, we obtain
\begin{align}
&\bra{\xi(\bar k\eta+\beta+u)}\mathcal K_{\rm R}\left[\ket{\psi(\bar k\eta+\beta)}\bra{\psi(\bar k'\eta+\beta)}\right]\ket{\xi(\bar k'\eta+\beta+u)}\no\\
&=\frac{\bell{1}(u)\bell{2}(\frac{\delta+\eta+\bar k\eta+\beta+u}{2})}{\kappa\,\bell{1}(\eta)\bell{2}(\frac{\delta-\eta+\bar k\eta+\beta+u}{2})}\left[\frac{\bell{1}(u)\bell{2}(\frac{\delta+\eta+\bar k'\eta+\beta+u}{2})}{\kappa\,\bell{1}(\eta)\bell{2}(\frac{\delta-\eta+\bar k'\eta+\beta+u}{2})}\right]^*\frac{\braket{\psi(\bar k'\eta+\beta-\eta)}{\psi(\bar k\eta+\beta-\eta)}}{\braket{\psi(\delta)}{\psi(\delta)}}\no\\
&\quad \times \braket{\xi(\bar k\eta+\beta+u)}{\psi(\delta-\eta)}\braket{\psi(\delta-\eta)}{\xi(\bar k'\eta+\beta+u)}\no\\
&=-\frac{\bell{1}^2(u)\bell{1}\left(\nu+k\eta\right)\bell{1}(\nu^*+k'\eta) \bell{2}(\Ga-k\eta+\nu+u) \bell{2}(\Ga^*-k'\eta+\nu^*-u) }{\bell{1}(\eta-u) \bell{1}(\eta+u)}\no\\
&\quad \times \frac{\bell{3}(i \Ga _1-(k-k')\eta) \bell{4}(\Ga_0-(k+k'+2)\eta)}{\bell{3}(i\Ga_1+\nu-\nu^*+u) \bell{4}(\Ga_0+\nu+\nu^*)}.\label{R:RHS:exp:2}
\end{align}
From Eqs. \eqref{R:LHS:exp}, \eqref{R:RHS:exp:1} and \eqref{R:RHS:exp:2}, we can easily conclude that
\begin{align*}
\bra{\xi(\bar k\eta+\beta+u)}{{\rm LHS}}\ket{\xi(\bar k'\eta+\beta+u)}=\bra{\xi(\bar k\eta+\beta+u)}{{\rm RHS}}\ket{\xi(\bar k'\eta+\beta+u)}.
\end{align*}
Furthermore, by repeating the above procedure, one can obtain
\begin{align}
\bra{\xi(\bar k\eta+\beta+\epsilon u)}{{\rm LHS}}\ket{\xi(\bar k'\eta+\beta+\epsilon'u)}=\bra{\xi(\bar k\eta+\beta+\epsilon u)}{{\rm RHS}}\ket{\xi(\bar k'\eta+\beta+\epsilon 'u)},\quad \epsilon,\epsilon'=\pm 1.\label{LHS:RHS}
\end{align}
Since both sides of \eqref{eq(kkpR)} are 
$2\times 2$ matrices, the four equations presented in Eq. \eqref{LHS:RHS} consequently lead to \eqref{eq(kkpR)}.

\paragraph*{Case (A)}

Now we consider the case (A). Multiplying Eq. \eqref{R:LHS} from the left by $\bra{\xi(\bar k\eta+\beta+u)}$ and from the right by $\ket{\xi(\bar k'\eta+\beta+u)}$ and using Eqs. \eqref{theta:eq:1}, \eqref{theta:eq:3}, the LHS reads
\begin{align}
&\bra{\xi(\bar k\eta+\beta+u)}{{\rm LHS}}\ket{\xi(\bar k'\eta+\beta+u)}\no\\
&=\frac{r_{k+1,k'+1}}{r_{k,k'}}\braket{\xi(-\bar k\eta-\be)}{\xi(-\bar k'\eta-\be)}\bra{\xi(\bar k\eta+\beta+u)}\ket{\psi(\bar k\eta-u+\be)}\bra{\psi(\bar k'\eta-u+\be)}\ket{\xi(\bar k'\eta+\beta+u)}\no\\
&=\frac{\bell{1}^2(u) \bell{1}(\nu+k\eta) \bell{1}(\nu^*-k'\eta)\bell{2}(\Ga-(2k+1)\eta  )\bell{2}(\Ga^*+(2k'+1)\eta)\bell{2}(\Ga- k\eta+\nu+u) \bell{2}(\Ga^*+ k'\eta+\nu^*+u) }
{\bell{1}(u+(k+1)\eta+\nu)\bell{1}(u-(k'+1)\eta+\nu^*)\bell{2}(\Ga-(k+1)\eta+\nu) \bell{2}(\Ga^*+(k'+1)\eta+\nu^*)}\no\\
&\quad\times \frac{\bell{3}(i \Ga_1-(k+k'+1)\eta ) \bell{3}(i\Ga_1-(k+k'+2)\eta) \bell{4}(\Ga_0-(k-k')\eta)}{\bell{3}(i\Ga_1-(k+k')\eta)}.\label{R:LHS:exp:A}
\end{align}
The c-term in Eq. \eqref{R:RHS}, aappearing in the first and second lines of the last step, can be simplified as
\begin{align}
&\quad \braket{\xi(-u-\bar k\eta-\beta)}{\xi(-u-\bar k'\eta-\beta)}+\frac{r_{k,k'+1}}{r_{k,k'}}\braket{\xi(-u-\bar k\eta-\beta)}{\xi(u-\bar k'\eta-\beta)}\no\\
&\quad+\frac{r_{k+1,k'}}{r_{k,k'}}\braket{\xi(u-\bar k\eta-\beta)}{\xi(-u-\bar k'\eta-\beta)}+\frac{r_{k+1,k'+1}}{r_{k,k'}}\braket{\xi(u-\bar k\eta-\beta)}{\xi(u-\bar k'\eta-\beta)}\no\\
&\overset{\eqref{theta:eq:1}}{=}\bell{3}(i\Ga_1-(k+k'+1)\eta)\bell{4}(u+\Ga_0-(k-k')\eta)+\frac{r_{k,k'+1}}{r_{k,k'}}\bell{3}(u+i\Ga_1-(k+k'+1)\eta)\bell{4}(\Ga_0-(k-k')\eta)\no\\
&\quad+\frac{r_{k+1,k'}}{r_{k,k'}}\bell{3}(u-i\Ga_1+(k+k'+1)\eta)\bell{4}(\Ga_0-(k-k')\eta)\no\\
&\quad+\frac{r_{k+1,k'+1}}{r_{k,k'}}\bell{3}(i\Ga_1-(k+k'+1)\eta)\bell{4}(u-\Ga_0+(k-k')\eta)\no\\
&\overset{\eqref{theta:eq:2}}{=}\frac{\bell{1}(u-\eta ) \bell{1}(u+\eta) \bell{2}(\Gamma -(2k+1)\eta) \bell{2}(\Ga^*+(2 k'+1)\eta)}{\bell{1}(u+(k+1)\eta+\nu) \bell{1}(u-(k'+1)\eta+\nu^*)\bell{2}(\Gamma-  (k+1)\eta+\nu ) \bell{2}(\Ga^*+(k'+1)\eta+\nu^*) }\no\\
&\quad \times \frac{\bell{3}(i \Ga_1+\nu -\nu^*) \bell{3}(i \Ga_1-(k+k'+1)\eta) \bell{4}(\Ga_0+\nu +\nu^*+u)}{\bell{3}(i\Ga_1-(k+k')\eta) }.\label{R:RHS:exp:1A}
\end{align}
Multiplying the third line in the last step of Eq. \eqref{R:RHS} on the left by $\bra{\xi(\bar k\eta+\beta+u)}$ and on the right by $\ket{\xi(\bar k'\eta+\beta+u)}$, and applying Eqs. \eqref{theta:eq:1}, \eqref{theta:eq:3} and \eqref{U:psi:5}, we arrive at
\begin{align}
&\bra{\xi(\bar k\eta+\beta+u)}\mathcal K_{\rm R}\left[\ket{\psi(\bar k\eta+\beta)}\bra{\psi(\bar k'\eta+\beta)}\right]\ket{\xi(\bar k'\eta+\beta+u)}\no\\
&=\frac{\bell{1}(u)\bell{2}(\frac{\delta+\eta+\bar k\eta+\beta+u}{2})}{\kappa\,\bell{1}(\eta)\bell{2}(\frac{\delta-\eta+\bar k\eta+\beta+u}{2})}\left[\frac{\bell{1}(u)\bell{2}(\frac{\delta+\eta+\bar k'\eta+\beta+u}{2})}{\kappa\,\bell{1}(\eta)\bell{2}(\frac{\delta-\eta+\bar k'\eta+\beta+u}{2})}\right]^*\frac{\braket{\psi(\bar k'\eta+\beta-\eta)}{\psi(\bar k\eta+\beta-\eta)}}{\braket{\psi(\delta)}{\psi(\delta)}}\no\\
&\quad \times \braket{\xi(\bar k\eta+\beta+u)}{\psi(\delta-\eta)}\braket{\psi(\delta-\eta)}{\xi(\bar k'\eta+\beta+u)}\no\\
&=\frac{\bell{1}^2(u)\bell{1}(\nu+k\eta)\bell{1}(\nu^*-k'\eta)\bell{2}(\Ga-k\eta+\nu+u)\bell{2}(\Ga^*+k'\eta+\nu^*+u)}{\bell{1}(u-\eta)\bell{1}
(u+\eta)}\no\\
&\quad \times \frac{\bell{3}(i\Ga_1-(k+k'+2)\eta)\bell{4}(\Ga_0-(k-k')\eta)}{\bell{3}(i\Ga_1+\nu-\nu^*)\bell{4}(\Ga_0+\nu+\nu^*+u)}.\label{R:RHS:exp:2A}
\end{align}
Comparing Eqs. \eqref{R:LHS:exp:A} with \eqref{R:RHS:exp:1A} and \eqref{R:RHS:exp:2A}, we find
\begin{align*}
\bra{\xi(\bar k\eta+\beta+u)}{{\rm LHS}}\ket{\xi(\bar k'\eta+\beta+u)}=\bra{\xi(\bar k\eta+\beta+u)}{{\rm RHS}}\ket{\xi(\bar k'\eta+\beta+u)}.
\end{align*}
Analogously, by repeating the above procedure, one can verify
\begin{align*}
\bra{\xi(\bar k\eta+\beta+\epsilon u)}{{\rm LHS}}\ket{\xi(\bar k'\eta+\beta+\epsilon'u)}=\bra{\xi(\bar k\eta+\beta+\epsilon u)}{{\rm RHS}}\ket{\xi(\bar k'\eta+\beta+\epsilon 'u)},\quad \epsilon,\epsilon'=\pm 1.
\end{align*}
In this way, the correctness of Eq. \eqref{eq(kkpR)} in case (A) is also proved.

\end{widetext}

\end{document}